\documentclass[aps,showpacs,dvips,showkeys,preprintnumbers]{revtex4}

\usepackage{graphicx}
\usepackage{color}
\usepackage{subfigure}
\usepackage{amsmath}

\def \be  {\begin{equation}}
\def \ee  {\end{equation}}
\def \ee  {\end{equation}}
\def \bea {\begin{eqnarray}}
\def \eea {\end{eqnarray}}

\begin{document}

\preprint{ECTP-2019-01}
\preprint{WLCAPP-2019-01}
\vspace*{3mm}

\title{Almost-Entirely Empirical Estimation for Chemical Potential}

\author{Abdel Nasser Tawfik}
\email{atawfik@nu.edu.eg}
\affiliation{Nile University - Egyptian Center for Theoretical Physics, Juhayna Square off 26th-July-Corridor, 12588 Giza, Egypt}
\affiliation{Institute for Theoretical Physics, Goethe University, Max-von-Laue-Str. 1, D-60438 Frankfurt am Main, Germany}

\author{Magda Abdel Wahab}
\affiliation{Physics Department, Faculty of Women for Arts, Science and Education, Ain Shams University, 11577 Cairo, Egypt}

\author{Hayam Yassin}
\affiliation{Physics Department, Faculty of Women for Arts, Science and Education, Ain Shams University, 11577 Cairo, Egypt}

\author{Hadeer M. Nasr El Din}
\affiliation{Basic Science Department, Modern Academy for Engineering, 11571 Cairo, Egypt}

\begin{abstract}

Based on statistical thermal approaches, the transverse momentum distribution of the well-identified produced particles,  $\pi^+$, $\pi^-$, $K^+$, $K^-$, $p$, $\bar{p}$, is studied. From the partition function of grand-canonical ensemble, we propose a generic expression for the dependence of the full chemical potential $\mu$ on rapidity $y$. Then, by fitting this expression with the experimental results of most central $p_{\perp}$ and $d^2 N/2 \pi p_{\bot} dp_{\bot} dy$, at $7.7$, $11.5$, $19.6$, $27$, $39$, $130$, $200~$GeV, we introduce a generic expression for the rapidity dependence of $\mu$, at different energies and particle types, $\mu=a+b y^2$. The resulting energy dependence reads $\sqrt{s_{\mathtt{NN}}}=c[(\mu-a)/b]^{d/2}$. As a validation check, the proposed approach reproduces, excellently, the rapidity spectra measured at different energies.

\end{abstract}

\pacs{25.75.Dw, 74.62.-c, 25.75.-q}
\keywords{Particle production in relativistic collisions, transition temperature variations (phase diagram), Relativistic heavy-ion collisions, Relativistic heavy-ion collisions}

\maketitle

\section{Introduction}

With the relativistic heavy-ion experiments, it is intended - among others - to explore the various stages of the  most-central heavy-ion collision as an experimentally controlled attempt towards a better understanding of the collision dynamics and the subsequent evolution of the strongly colliding system \cite{Back:2002uc}. The deconfinement of colliding hadrons which are conjectured to expand rapidly and later on to cool down leads to formation of colored quark-gluon plasma (QGP), at high temperatures \cite{Ahmad:2013una}. Second, the "hadronization" or the phase transition from the deconfined QGP to the confined colorless hadrons can be simply illustrated as a recombination of quarks and gluons that takes place at some critical temperatures. After cooling down, a stage characterized by fixing the number of uncorrelated produced particles, i.e. chemical freezeout, takes place \cite{Tawfik:2016jzk,Tawfik:2013eua,Tawfik:2013dba,Tawfik:2012si,Tawfik:2005qn,Tawfik:2004ss,Tawfik:2004vv}. In such an expanding system, another freezeout state is to be defined by a competition between local scattering and the expansion rates, known as kinetic freezeout \cite{Zimanyi1978}.

Various signatures of the QGP formation and the phase transition have been suggested, so far \cite{Tawfik:2014eba}. One of these signatures is the strangeness enhancement where the gluon-gluon interactions in  QGP-medium lead to $s\bar{s}$ pair-production and this saturates rapidly the content of strange quark in case of forming QGP phase resulting in an enhancement in the production of strange hadrons and hence the multi-strange baryons and strange anti-baryons, as well \cite{Letessier:1993qa,ALICE:2017jyt}. Another type of signature is that the various thermodynamic quantities such as pressure, energy density, and entropy density of the system can be estimated statisticaly as functions of the temperature $T$ and the baryon chemical potential $\mu_{\mathtt{B}}$. A third type of indirect signatures, which manifests essential properties of QGP, can be measured by the transverse momentum distribution $p_{\mathtt{T}}$ \cite{Cleymans:1998fq,Magas:2003wi}. These are examples on the significant tools available in order to understand the dynamics and the properties of the produced particles from initial stages of the collision up to the final freezeout stage is the precise measurement of the $p_{\mathtt{T}}$ of well-identified hadrons with varying rapidity spectra \cite{Uddin:2011bi}. In the present work, we combine the second and third types of signatures. We aiming at an empirical estimation of the chemical potential $\mu$ from the rapidity distributions of well-identified produced particles, $\pi^+$, $\pi^-$, $K^+$, $K^-$, $p$, $\bar{p}$.

The dependence of the freezeout temperature $T_{\mathtt{ch}}$ on the fireball rapidity was discussed in literature \cite{Becattini:2007ci,Becattini:2007qr,Cleymans:2008zz,Cleymans:2007jj}. Another concept for the dependence of the baryon chemical potential $\mu_{\mathtt{B}}$ on the rapidity $y$ was reported in ref. \cite{Uddin:2010}. It was concluded that a certain value of temperature seems being sufficient to describe the results obtained at different rapidity intervals. In principle, the chemical potential $\mu$ plays an essential role in the statistical treatment of the particle production. 

The rapidity spectra of protons and anti-protons measured in Relativistic Heavy Ion Collider (RHIC) at nucleon-nucleon center-of-mass energy $\sqrt{s_{\mathtt{NN}}}=200~$GeV have been described by a statistical model \cite{Becattini:2007qr}, in which the formation of hot and dense regions was assumed to move along the beam axis with increasing rapidity. An extended version of this model was utilized in describing the rapidity spectra of protons, anti-protons, Kaons, anti-Kaons, pions, and also the ratios $\bar{\Lambda}/\Lambda$ and $\bar{\Xi}/\Xi$ measured at RHIC energies \cite{Uddin:2009df,Uddin:2009zi}. Out of these proposals a unified approach was constructed \cite{Uddin:2011bi}.

In the present paper, we introduce an alternative method to determine - in an almost direct way - the chemical potential $\mu$ from the rapidity distributions of well-identified produced particles, $\pi^+$, $\pi^-$, $K^+$, $K^-$, $p$, $\bar{p}$. In almost all high-energy experiments, the rapidity distributions are precisely measured quantities. This means that when utilizing this measured quantity, we can quasi directly access or determine the chemical potentials of the various particles produced within a definite range of rapidity, independent on the collision energy, or the collision volume and/or its type. In other words, we are introducing a universal approach enabling to {\it measure} the chemical potentials in an almost-entirely empirical procedure. The ultimate confirmation of our results, i.e. proposing a universal dependence of the full chemical potentials $\mu$ on rapidity of each produced particle, could be proven through the ability to reproduce the measured rapidity distributions for these particles. In other words, should our proposal for the dependence of $\mu$ on $y$ is valid, measured $dN/dy$ vs. $y$ in most-central collisions can well be described, at various collision energies.

The present paper is organized as follows. The theoretical approaches are presented in section \ref{sec:theor}. The statistical thermal approaches are discussed in section \ref{subsec:thermal}. Section \ref{sec:res} gives the results. The dependence of the resulting chemical potential on rapidity is shown in section \ref{subsec:mu_y}. The utilization of our approach in reproducing various experimentally measured rapidity spectra, $dN/dy$ vs. $y$, throughout the proposed approach is to be verified, shall be discussed in section \ref{Impl}. The final conclusions are outlined in section \ref{sec:cncl}

\section{The Model}
\label{sec:theor}

The momentum distribution of particle emerging from a fireball can be expressed as \cite{Rafelski_Letessier:2004,Becattini:2007ci,Uddin:2009df}
\begin{eqnarray}
E \frac{d^3 N}{d^3 p}\equiv f(E,p_L),
\label{eq:1}
\end{eqnarray}
where $E$ and $p_L$ are the energy and the longitudinal momentum of such a particle, respectively. The momentum-space volume element reads
\begin{eqnarray}
\frac{d^3 p}{E} = m_\perp dm_\perp dy d\varphi.
\label{eq:2}
\end{eqnarray}
Then, in terms of the transverse mass, the particle distribution can be obtained 
\begin{eqnarray}
\frac{d^3 N}{dy m_\perp dm_\perp d\varphi} \equiv  f(E,p_L), \qquad & \mathtt{or} & \qquad
\frac{d^2 N}{2 \pi m_\perp dm_\perp dy} \equiv  f(E,p_L).
\label{eq:3}
\end{eqnarray}
At mid-rapidity, i.e. $y=0$, the transverse momentum distribution deduced from Eq. (\ref{eq:1}) reads
\begin{eqnarray}
\frac{d^2 N}{2 \pi p_{\perp} dp_{\perp} dy}&\equiv & f(E,p_L).
\label{eq:4}
\end{eqnarray}
In the section that follows, we introduce our approach, which aims at an almost empirical determination of the  chemical potential $\mu$ from measured rapidity distribution.

\subsection{Our Approach}
\label{subsec:thermal}

In order to characterize the particle production in the final state, where the chemical freezeout is likely reached, at which the number of produced particles is fixed in the end, different statistical-thermal approaches have been utilized. For a recent review, interested reads are advised to consult ref. \cite{Tawfik:2014eba}. The expressions outlined in the previous section can be now formulated within a statistical-thermal model, such as the hadron resonance gas model (HRGM) \cite{Tawfik:2014eba}. For simplicity let us assume that the HRGM has just one constituent,
\begin{eqnarray}
Z(T, \mu, V) &=& \mathtt{Tr} \left[ \exp\left(\frac{\mu\,N-H}{T}\right) \right], \label{eq:lnZ}
\end{eqnarray}
where $H$ is the Hamiltonian of the system, which is given as a summation of the kinetic energies of the relativistic Fermi and Bose constituents counting for the effective degrees-of-freedom of the confined hadrons \cite{Tawfik:2004sw}. 

The chemical potential $\mu$ combines all components related to the various quantum numbers; $\mu=B\mu_{\mathtt{B}}+S\mu_{\mathtt{S}}+Q\mu_{\mathtt{Q}}+I_3\mu_{\mathtt{I_3}}+\cdots$, where $B$, $S$, $Q$, and $I_3$ are baryon, strange, electric charge, and isospin quantum numbers, respectively. 

The freezeout temperature $T_{\mathtt{ch}}$ and  $\mu$ are well-known thermodynamic parameters to be determined from these approaches \cite{Tawfik:2012zp}. In Fermi-Dirac and Bose-Einstein statistics, the total number of particles $N$ can be deduced as
\begin{equation}
N = \frac{g V}{(2 \pi)^3} \int \left(\exp\left[ \frac{\mu - E}{T}\right] \pm 1 \right)^{-1} d^3p,
\label{numdens}
\end{equation}
where $g$ is the degeneracy factor and $V$ is the volume of the system of interest. $\pm$ stands for fermions and bosons, respectively. From Eq. (\ref{numdens}), the momentum distribution, Eqs. (\ref{eq:2})-(\ref{eq:4}), can be estimated
\begin{eqnarray}
E \frac{d^3 N}{d^3p}=\pm E\frac{g V}{(2 \pi)^3}\left(\exp\left[ \frac{\mu-E}{T}\right] \pm1\right) ^{-1}. \label{eq:6}
\end{eqnarray}
From Eqs. (\ref{eq:4}) and (\ref{eq:6}), we get
\begin{eqnarray}
\frac{d^2 N}{2 \pi p_{\perp} dp_{\perp} dy}=\pm E\frac{g V}{(2 \pi)^3}\left(\exp\left[ \frac{\mu-E}{T}\right] \pm1\right)^{-1}.
\label{eq:7}
\end{eqnarray}
Hence energy can be expressed as a function of rapidity, $E=m_\bot \cosh(y)$ \cite{Rafelski_Letessier:2004} and the transverse mass of the particle $m_{\bot}$ can be given in dependence on the transverse momentum of that particle $p_{\bot}$ and on its mass $m_{\bot}=(m^2+p_{\bot}^2)^{1/2}$, then
\begin{eqnarray}
\frac{d^2 N}{2 \pi p_{\bot} dp_{\bot} dy} &=& \pm m_{\bot} \cosh(y) \frac{g V}{(2 \pi)^3} \left(\exp\left[\frac{\mu - m_\bot \cosh(y)}{T}\right] \pm 1 \right)^{-1}. \label{eq:8}
\end{eqnarray} 
From Eq. (\ref{eq:8}), we can deduce an expression for the dependence of the total chemical potential $\mu$ of the particle of interest on its measured rapidity $y$,
\begin{eqnarray}
\mu = m_\bot \cosh(y) - T \ln \left[\frac{\pm  \frac{g V}{(2 \pi)^3} m_\bot \cosh(y)}{\frac{d^2 N}{2 \pi p_{\bot} dp_{\bot} dy}} \mp 1\right].
\label{eq:9}
\end{eqnarray}
In other words,  we propose to express $\mu$ in terms of $y$ besides $m_\bot$ and $d^2 N/(2 \pi p_{\bot} dp_{\bot} dy)~$GeV$^{-2}$. The latter is a quantity, which is precisely measured in most-central collisions, Tabs. \ref{Tab:1}-\ref{Tab:6}. The energy dependence of $\mu$ is hidden in $y$ and apparently in $p_{\perp}$ and  $d^2 N/(2 \pi p_{\bot} dp_{\bot} dy)$, as well. Eq. (\ref{eq:pi+mu}) translates this in an unambiguous way. 

Also, by integrating Eq. (\ref{eq:8}) with respect to $p_{\bot}$, the particle rapidity distribution $dN/dy$ can be given as a function of rapidity $y$ 
\begin{eqnarray}
\frac{dN}{dy} = \pm \frac{g V}{(2 \pi)^3} \int \frac{\pm \sqrt{m^2+p_{\bot}^2} \cosh(y) p_{\bot}}{\left(\exp\left[\frac{\mu-m_\bot \cosh(y)}{T} \right] \pm 1\right)} dp_{\bot}.
\label{eq:10}
\end{eqnarray}
This is another expression, with which we can judge about the validity of our approach, where $\mu$ given in Eq. (\ref{eq:9}) plays an essential role. Again, it is apparent that apart from $\mu$, both $p_{\bot}$ and $m_{\bot}$ are characterizing the particle of interest. Accordingly, we believe that the physical insights of the dependence $dN/dy$ on $y$ can be learned from Eq.  (\ref{eq:10}). The latter is a precise measurement. Should our approach for $\mu$ is correct, Eq. (\ref{eq:10}) should be able to reproduce the experimental results. The energy dependence of $dN/dy$ is also given in $y$.

\section{Results and discussion}
\label{sec:res}

For the seek of completeness, we recall that the dependence of the baryon chemical potential on the beam energies was subject of various studies \cite{Tawfik:2013bza,Andronic:2005yp}. In the present study, we focus on an almost entire empirical estimation of the dependence of the {\it generic} chemical potential on rapidity of well-identified produced particles. With {\it generic} we mean all quantum numbers, especially baryon and strangeness. While the energy dependence reported in refs. \cite{Tawfik:2013bza,Andronic:2005yp}  is based on thermal models and their confrontation to the experimental results from most-central collisions, none of such models is needed for the rapidity dependence we are introducing here. The second difference is that the thermal model energy dependence \cite{Tawfik:2013bza,Andronic:2005yp} deals with the baryon chemical potential, exclusively. The {\it generic} chemical potential introduced in the present script is analyzed in dependence on rapidity of well-identified produced particles, only. The third difference is the thermodynamic nature of both quantities. While $\mu$ is intensive, the energy is an extensive thermodynamic quantity. 

\subsection{rapidity dependence of resulting chemical potential}
\label{subsec:mu_y}

\begin{figure}[!htb]
\includegraphics[width=5.cm]{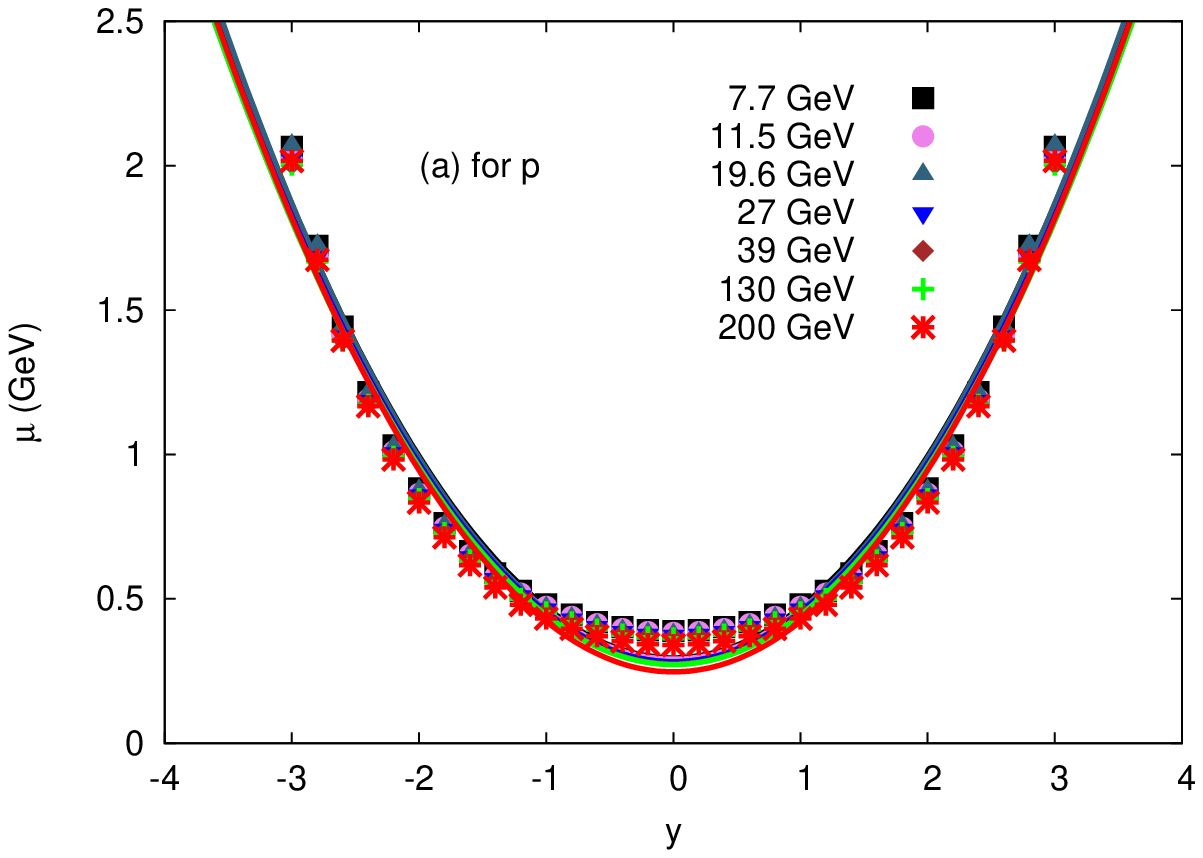}
\includegraphics[width=5.cm]{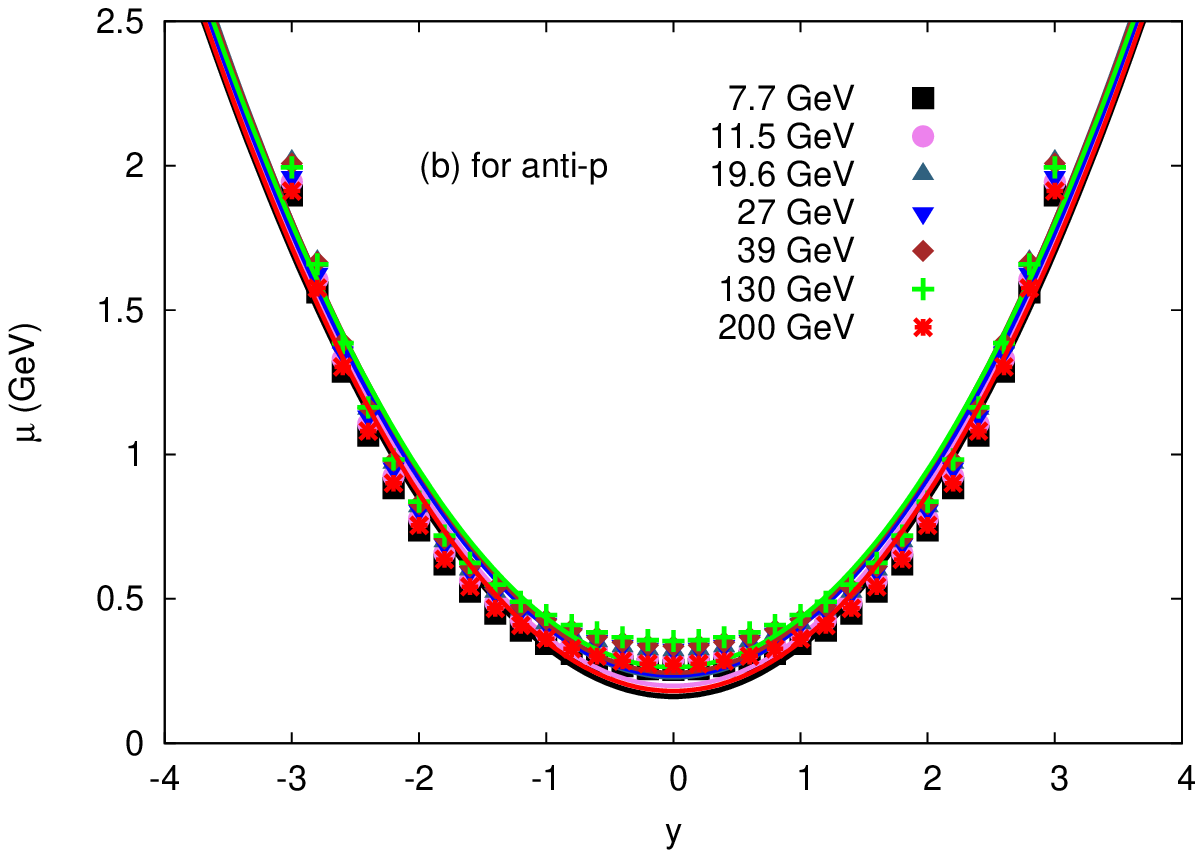} 
\includegraphics[width=5.cm]{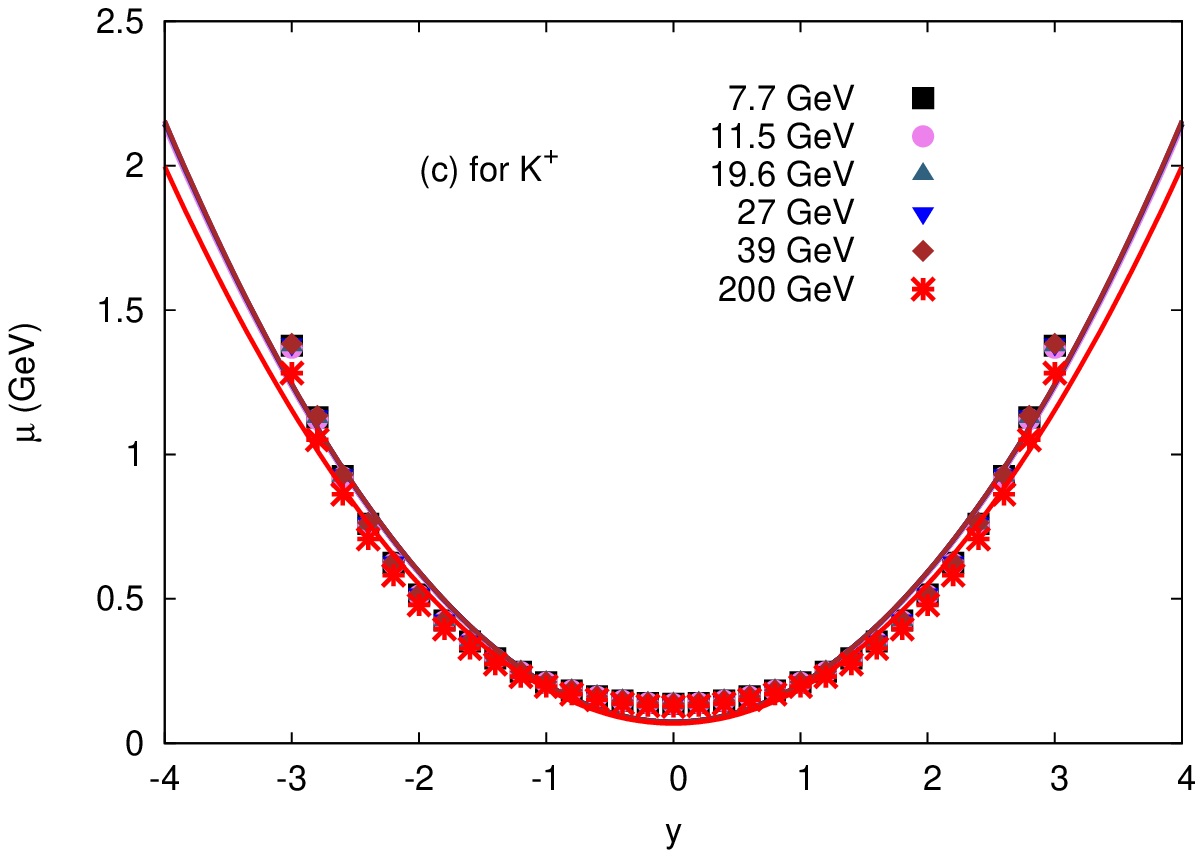}
\includegraphics[width=5.cm]{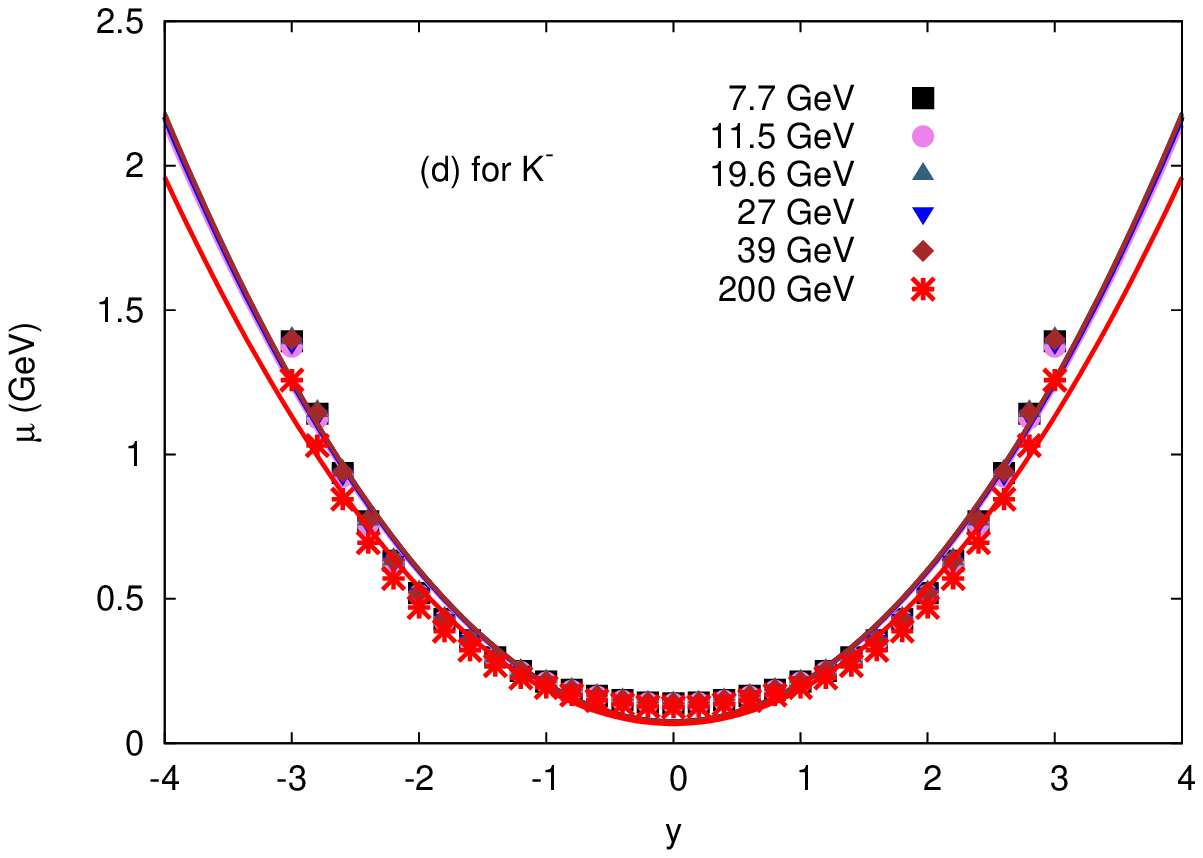} 
\includegraphics[width=5.cm]{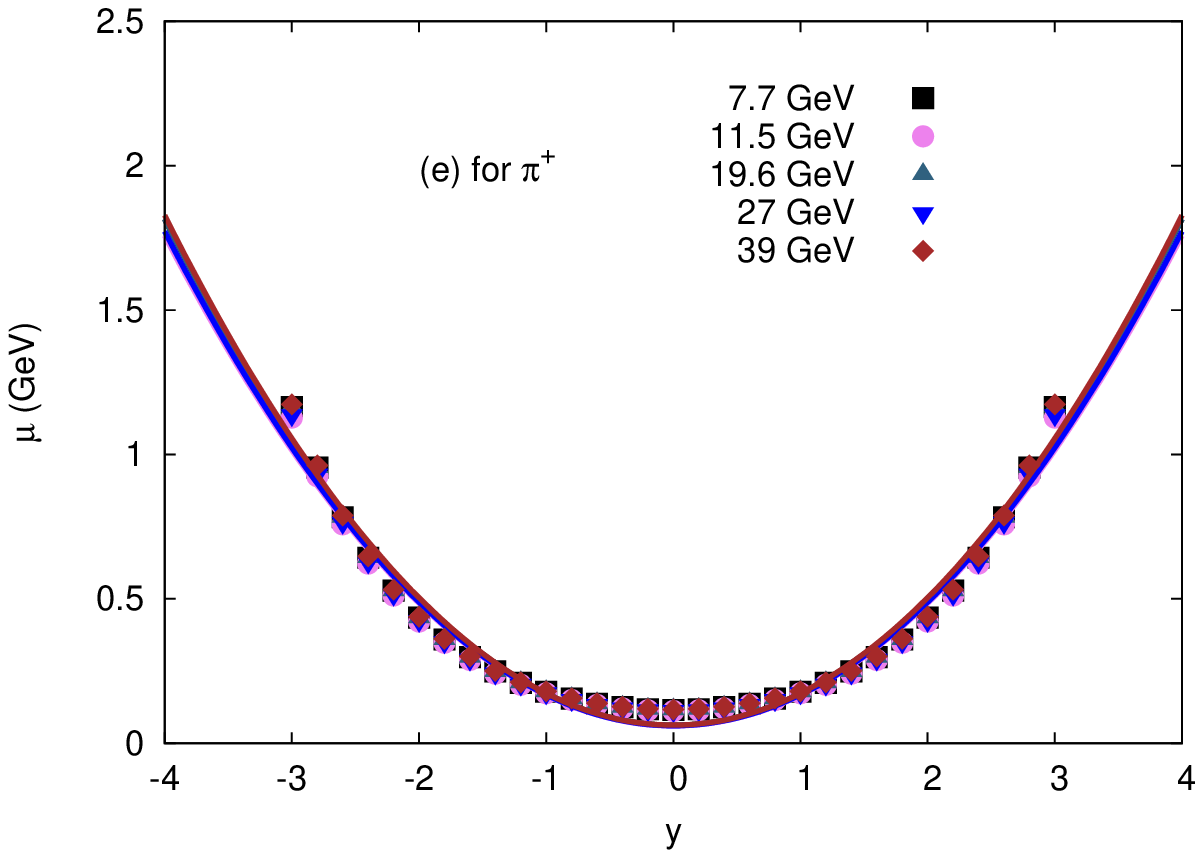}
\includegraphics[width=5.cm]{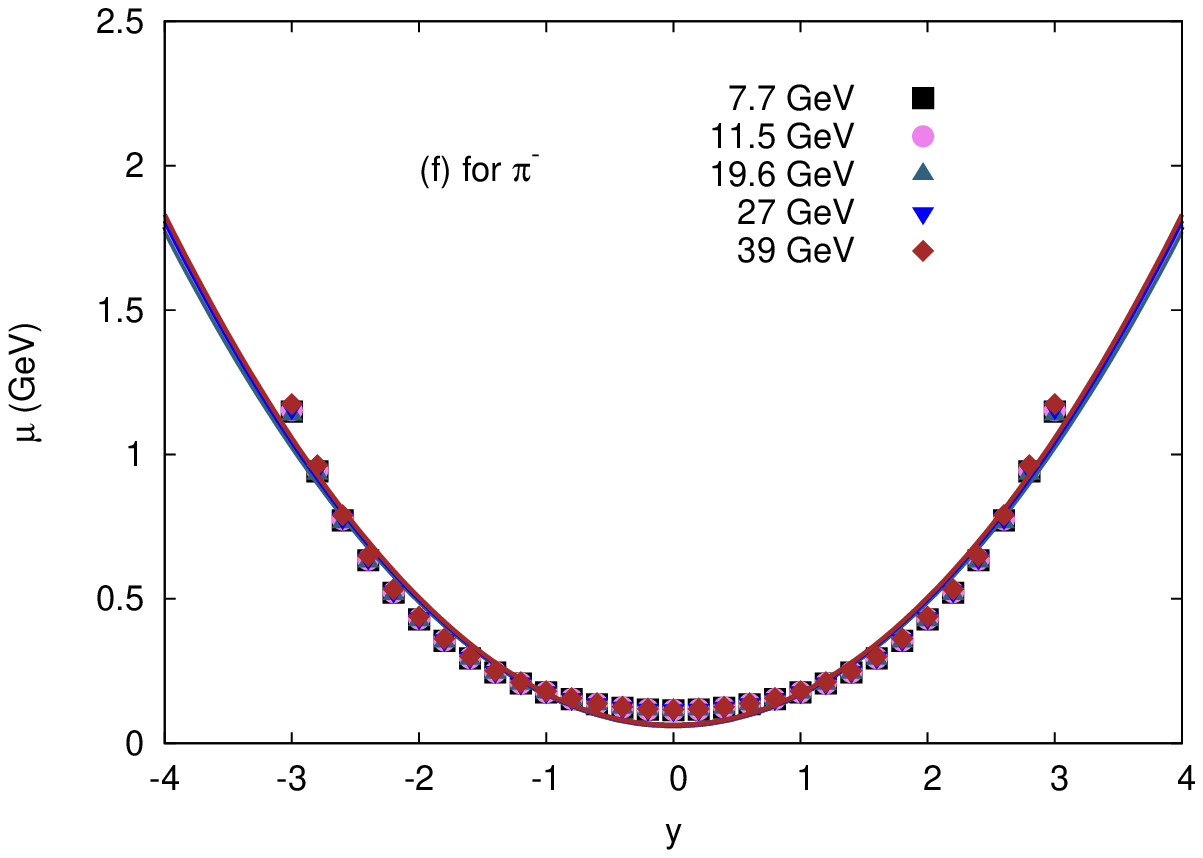}
\caption{Results of the chemical potential $\mu$ as functions of rapidity $y$ for $\pi^+$, $\pi^-$, $K^+$, $K^-$, $p$, $\bar{p}$ are depicted in panels (a), (b), (c), (d), (e), and (f), respectively. Symbols refer to the calculations based on Eq. (\ref{eq:9}), where STAR results on $p_{\perp}$ in GeV and $d^2 N/(2 \pi p_{\perp} dp_{\perp} dy)~$GeV$^{-2}$, at $\sqrt{s_{\mathtt{NN}}}=7.7$, $11.5$, $19.6$, $27$, $39$, $130$, $200~$GeV \cite{Adamczyk:2017iwn} are taken into account, while the curves represent the statistical fits, Eqs. (\ref{eq:pi+mu})-(\ref{eq:apmu}) and Tabs. \ref{Tab:1}-\ref{Tab:6}. \label{fig:1}}
\end{figure}

The dependence of the chemical potential $\mu$ on rapidity $y$ is (can be) estimated from the partition function of the grand canonical ensemble \cite{Tawfik:2014eba}. As discussed, the proposed expression, Eq. (\ref{eq:9}), is deduced from the integration of Eq. (\ref{eq:7}). Then, for various charged particles at different energies, the values obtained from this expression are to be inserted in all forthcoming equations up to Eq. (\ref{eq:9}). As elaborated in the previous section, Eq. (\ref{eq:9}) can be estimated by using experimental inputs for $p_{\perp}$ in GeV and  $d^2 N/(2 \pi p_{\bot} dp_{\bot} dy)~$GeV$^{-2}$. We assume an over all freezeout temperature $T=0.165~$GeV at various  collision energies, as shown in Fig. \ref{fig:1}. With this respect, it should be emphasized that to the authors' best knowledge there is no experimental results available on rapidity distribution at $64~$GeV to compare with. 

Alternatively, from Eq. \ref{eq:9}, a direct relation between $\mu$ and $y$ is proposed. This is the main target of the present script. To this end, we first fit this expression to the experimental results from most-central collisions on $p_{\perp}$ in GeV and $d^2 N/(2 \pi p_{\bot} dp_{\bot} dy)$ in GeV$^{-2}$ for $\pi^+$, $\pi^-$ $p$, $K^+$, $K^-$, $\bar{p}$, at $\sqrt{s_{\mathtt{NN}}}=7.7$, $11.5$, $19.6$, $27$, $39$, $130$, $200~$GeV \cite{Adamczyk:2017iwn}, Fig. \ref{fig:1}. Fig. \ref{fig:1} depicts $\mu$ as function of $y$. The symbols are the calculations based on Eq. (\ref{eq:9}), where $p_{\perp}$ in GeV and $d^2 N/(2 \pi p_{\bot} dp_{\bot} dy)$ in GeV$^{-2}$ for $\pi^+$, $\pi^-$, $K^+$, $K^-$, $p$, $\bar{p}$, at $\sqrt{s_{\mathtt{NN}}}=7.7$, $11.5$, $19.6$, $27$, $39$, $130$, $200~$GeV are taken from the STAR experiment \cite{Adamczyk:2017iwn}, Tabs. \ref{Tab:1}-\ref{Tab:6}, while the curves are the statistical fits based on the expressions (\ref{eq:pi+mu}) - (\ref{eq:apmu}). 

It is worthy emphasizing that ref. \cite{Adamczyk:2017iwn} isn't presenting experimental results on rapidity, explicitly. Indeed, it is merely introducing results on the transverse momentum distributions, $p_{\perp}$ in GeV and  $d^2 N/(2 \pi p_{\bot} dp_{\bot} dy)~$GeV$^{-2}$. These are the ones entering our calculations based on Eq. \ref{eq:9}. In other words, with the experimental results on the different produced particles, we mean that calculations based on the experimentally measured transverse momenta of these particles in most-central collisions. Again, we first use Eq. \ref{eq:9} to draw symbols in Fig. \ref{fig:1}. Then we fit these results to the various expressions (\ref{eq:pi+mu}) - (\ref{eq:apmu}). in order to obtain another dependence of $\mu$ on $y$ for various particles at different energies, Tabs. \ref{Tab:1}-\ref{Tab:6}. 

For the statistical fits, the values of $p_{\perp}$ and $d^2 N/(2 \pi p_{\perp} dp_{\perp} dy)~$GeV$^{-2}$ at different energies, Tabs. \ref{Tab:1}-\ref{Tab:6}, are taken into consideration. Then, the results of $\mu$ as functions of $y$ for the individual particles can be plotted, as in Fig. \ref{fig:1}. Again, the most-central experimental results on $p_{\perp}$ in GeV and $d^2 N/(2 \pi p_{\perp} dp_{\perp} dy)~$GeV$^{-2}$, make it possible to determine $\mu(y)$,  Eq. \ref{eq:9}.

\begin{table}[ht]
\centering 
\begin{tabular}{||c | c | c | c ||} \hline\hline
  $\sqrt{s_{\mathtt{NN}}}$ GeV & $p_{\perp}$ GeV & $\frac{d^2 N}{2 \pi p_{\perp} dp_{\perp} dy}~$GeV$^{-2}$ &  $\mu=a + b y^2$ (MeV)\\
   \hline
  7.7 & 0.6796 & 12.023 & $(0.062\pm0.014) + (0.109\pm0.003) y^2$ \\
  11.5 & 0.6573 & 15.811 & $(0.06\pm0.014) + (0.106\pm0.003) y^2$ \\
  19.6 & 0.68 & 21.796 & $(0.062\pm0.019) + (0.109\pm0.003) y^2$ \\
  27 & 0.663 & 24.669 & $(0.061\pm0.014) + (0.107\pm0.003) y^2$ \\
  39 & 0.685 & 25.724 & $(0.062\pm0.014) + (0.110\pm0.003) y^2$ \\
  \hline\hline 
  \end{tabular}
\caption{Experimental results on $p_{\perp}$ GeV and $d^2 N/(2 \pi p_{\perp} dp_{\perp} dy)~$GeV$^{-2}$, at various energies utilized in Eq. (\ref{eq:9}) and depicted in Fig. \ref{fig:1}. The very right column gives statistical fits for $\mu$ in dependence on $y$ and the fit parameters $a$ and $b$ for $m_{\pi^+}=0.140~$GeV and $g_{\pi^+}=1$. \label{Tab:1} }
\end{table}
\begin{table}[ht]
\centering 
\begin{tabular}{||c | c | c | c ||} \hline\hline
  $\sqrt{s_{\mathtt{NN}}}$ GeV & $p_{\perp}$ GeV &c $\frac{d^2 N}{2 \pi p_{\perp} dp_{\perp} dy}~$GeV$^{-2}$ &  $\mu=a + b y^2$ (MeV)\\
   \hline
  7.7 & 0.67 & 11.402 & $(0.061\pm0.014) + (0.108\pm0.003) y^2$ \\
  11.5 & 0.673 & 16.668 & $(0.061\pm0.014) + (0.108\pm0.003) y^2$ \\
  19.6 & 0.663 & 21.236 & $(0.061\pm0.014) + (0.107\pm0.003) y^2$ \\
  27 & 0.678 & 23.792 & $(0.062\pm0.014) + (0.109\pm0.003) y^2$ \\
  39 & 0.686 & 25.292 & $(0.063\pm0.014) + (0.110\pm0.003) y^2$ \\
  \hline\hline 
  \end{tabular}
\caption{The same as Tab. \ref{Tab:1} but for $m_{\pi^-}=0.140~$GeV and $g_{\pi^-}=1$. \label{Tab:2} }
\end{table}

\begin{table}[ht]
\centering 
\begin{tabular}{||c | c | c | c ||} \hline\hline
  $\sqrt{s_{\mathtt{NN}}}$ GeV  & $p_{\perp}$ GeV & $\frac{d^2 N}{2 \pi p_{\perp} dp_{\perp} dy} ~$GeV$^{-2}$ &  $\mu=a + b y^2$ (MeV)\\
   \hline
  7.7 & 0.658 & 5.014 & $(0.073\pm0.017) + (0.129\pm0.004) y^2$ \\
  11.5 & 0.653 & 5.464 & $(0.073\pm0.017) + (0.128\pm0.004) y^2$ \\
  19.6 & 0.658 & 6.552 & $(0.073\pm0.017) + (0.129\pm0.004) y^2$ \\
  27 & 0.662 & 7.536 & $(0.074\pm0.017) + (0.130\pm0.004) y^2$ \\
  39 & 0.663 & 6.888 & $(0.074\pm0.017) + (0.130\pm0.004) y^2$ \\
  200 & 0.586 & 11.911 & $(0.074\pm0.016) + (0.131\pm0.004) y^2$ \\
  \hline\hline 
  \end{tabular}
\caption{The same as Tab. \ref{Tab:1} but for $m_{K^+}=0.490~$GeV and $g_{K^+}=1$.  \label{Tab:3} }
\end{table}

\begin{table}[ht]
\centering 
\begin{tabular}{||c | c | c | c ||} \hline\hline
  $\sqrt{s_{\mathtt{NN}}}$ GeV & $p_{\perp}$ GeV & $\frac{d^2 N}{2 \pi p_{\perp} dp_{\perp} dy} ~$GeV$^{-2}$ &  $\mu=a + b y^2$ (MeV)\\
   \hline
  7.7 & 0.658 & 5.014 & $(0.073\pm0.017) + (0.129\pm0.0039) y^2$ \\
  11.5 & 0.653 & 5.464 & $(0.073\pm0.016) + (0.128\pm0.0034) y^2$ \\
  19.6 & 0.658 & 6.552 & $(0.073\pm0.017) + (0.129\pm0.004) y^2$ \\
  27 & 0.662 & 7.536 & $(0.073\pm0.017) + (0.129\pm0.0039) y^2$ \\
  39 & 0.663 & 6.888 & $(0.073\pm0.017) + (0.130\pm0.004) y^2$ \\
  200 & 0.586 & 11.911 & $(0.074\pm0.015) + (0.131\pm0.0035) y^2$ \\
  \hline\hline 
  \end{tabular}
\caption{The same as Tab. \ref{Tab:1} but for  $m_{K^-}=0.490~$GeV and $g_{K^-}=1$.  \label{Tab:4} }
\end{table}

\begin{table}[ht]
\centering 
\begin{tabular}{||c | c | c | c ||} \hline\hline
  $\sqrt{s_{\mathtt{NN}}}$ GeV & $p_{\perp}$ GeV & $\frac{d^2 N}{2 \pi p_{\perp} dp_{\perp} dy} ~$GeV$^{-2}$ &  $\mu=a + b y^2$ (MeV)\\
\hline
  7.7 & 0.668 & 11.926 & $(0.295\pm0.024) + (0.174\pm0.005) y^2$ \\
  11.5 & 0.622 & 10.778 & $(0.295\pm0.024) + (0.170\pm0.005) y^2$ \\
  19.6 & 0.687 & 6.983 & $(0.282\pm0.024) + (0.176\pm0.005) y^2$ \\
  27 & 0.634 & 7.679 & $(0.282\pm0.024) + (0.171\pm0.005) y^2$ \\
  39 & 0.626 & 5.835 & $(0.274\pm0.024) + (0.170\pm0.005) y^2$ \\
  130 & 0.625 & 5.39 & $(0.272\pm0.024) + (0.170\pm0.005) y^2$ \\
  200 & 0.667 & 2.038 & $(0.247\pm0.024) + (0.174\pm0.005) y^2$  \\ 
\hline\hline 
\end{tabular}
\caption{The same as Tab. \ref{Tab:1} but for $m_{p}=0.938~$GeV and $g_p=2$.  \label{Tab:5} }
\end{table}

\begin{table}[ht]
\centering 
\begin{tabular}{||c | c | c | c ||} \hline\hline
  $\sqrt{s_{\mathtt{NN}}}$ GeV  & $p_{\perp}$ GeV & $\frac{d^2 N}{2 \pi p_{\perp} dp_{\perp} dy} ~$GeV$^{-2}$ &  $\mu=a + b y^2$ (MeV)\\
   \hline
  7.7 & 0.63 & 0.098 & $(0.162\pm0.024) + (0.171\pm0.0055) y^2$ \\
  11.5 & 0.641 & 0.362 & $(0.198\pm0.024) + (0.172\pm0.0056) y^2$ \\
  19.6 & 0.687 & 0.944 & $(0.227\pm0.024) + (0.176\pm0.0057) y^2$ \\
  27 & 0.621 & 1.362 & $(0.234\pm0.024) + (0.170\pm0.0056) y^2$ \\
  39 & 0.662 & 1.803 & $(0.243\pm0.024) + (0.174\pm0.0057) y^2$ \\
  130 & 0.625 & 3.840 & $(0.263\pm0.024) + (0.171\pm0.005) y^2$ \\
  200 & 0.625 & 0.193 & $(0.180\pm0.024) + (0.171\pm0.005) y^2$ \\
  \hline\hline 
  \end{tabular}
\caption{The same as Tab. \ref{Tab:5} but for $m_{\bar{p}}=0.938~$GeV and $g_{\bar{p}}=2$.
 \label{Tab:6} }
\end{table}

The procedure of relating the chemical potential $\mu$ to the rapidity $y$ can be summarized as follows.
\begin{itemize}
\item Substituting the experimental values of $p_{\bot}~$GeV and $d^2 N/(2 \pi p_{\bot} dp_{\bot} dy)~$GeV$^{-2}$ from most-central collisions \cite{Adamczyk:2017iwn,Abelev:2008ab}, Tabs. \ref{Tab:1}-\ref{Tab:6}, in Eq. (\ref{eq:9}).
\item Then, plotting $\mu$ vs. $y$ as shown in Fig. \ref{fig:1}.
\item And finally extracting expressions for each of the produced particles, Tabs. \ref{Tab:1}-\ref{Tab:6}. 
\end{itemize}
This is the only fit process conducted in order to obtain generic expressions for $\mu(y)$.

It is obvious that just one expression is utilized for each particle and its anti-particle. It is obvious that this {\it universal} expression is not depending on the collision energies. We come again to this point. But as discussed in earlier sections, the energy dependence is hidden in $y$ among other quantities. When moving from one energy to another, we repeat the same procedure. Accordingly, the dependence of the resulting chemical potential of each particle $\mu$ on rapidity $y$ can be expressed as
\begin{itemize}
\item For $\pi^+$
\begin{equation} 
\mu=(0.06\pm0.001) + (0.109\pm0.01)\, y^2. \label{eq:pi+mu}
\end{equation}

\item For $\pi^-$
\begin{equation} 
\mu=(0.06\pm0.001) + (0.109\pm0.01)\, y^2. \label{eq:pi-mu}
\end{equation}

\item For $K^+$
\begin{equation} 
\mu=(0.07\pm0.001) + (0.13\pm0.001)\, y^2. \label{eq:k+mu}
\end{equation} 

\item For $K^-$
\begin{equation} 
\mu=(0.07\pm0.001) + (0.13\pm0.001)\, y^2. \label{eq:k-mu}
\end{equation} 

\item For $p$,  
\begin{equation} 
\mu=(0.28\pm0.001) + (0.17\pm0.001)\, y^2. \label{eq:pmu}
\end{equation}

\item For $\bar{p}$
\begin{equation} 
\mu=(0.28\pm0.001) + (0.17\pm0.001)\, y^2. \label{eq:apmu}
\end{equation}
\end{itemize} 
Obviously, this result would guide our thoughts that each particle would have a defined value of $\mu$ at a certain value of rapidity $y$ \cite{Becattini:2007ci,Uddin:2009df}. With these regards, it is worthy recalling again that each of these expressions seems not depending on the collision energy and being identical for the particle and its anti-particle. 

In light of this result, we would like to make a further step. We propose a more {\it universal} expression unifying the previous ones, i.e. convincingly relating $\mu$ to $y$ for all particles at all energies. 
\begin{eqnarray}
\mu = a + b\, y^2,  \label{eq:muy} 
\end{eqnarray}
where, $a$ and $b$ are parameters in MeV units slightly differ from the ones listed in Tabs. \ref{Tab:1}-\ref{Tab:6}, in Eq. (\ref{eq:9}).  This is the main result out of this study. In the section \ref{Impl}, we elaborate how this expression is valid, at all energies and well reproduces all experimental results for rapidity distributions, $dN/dy$ vs. $y$.  From Eq. (\ref{eq:muy}), we would conclude that vanishing rapidity might be accompanied by finite chemical potentials, $\mu=a$. 

\subsection{Energy dependence of resulting chemical potential}
\label{muEnrgy}

\begin{figure}[!htb]
\includegraphics[width=12.cm]{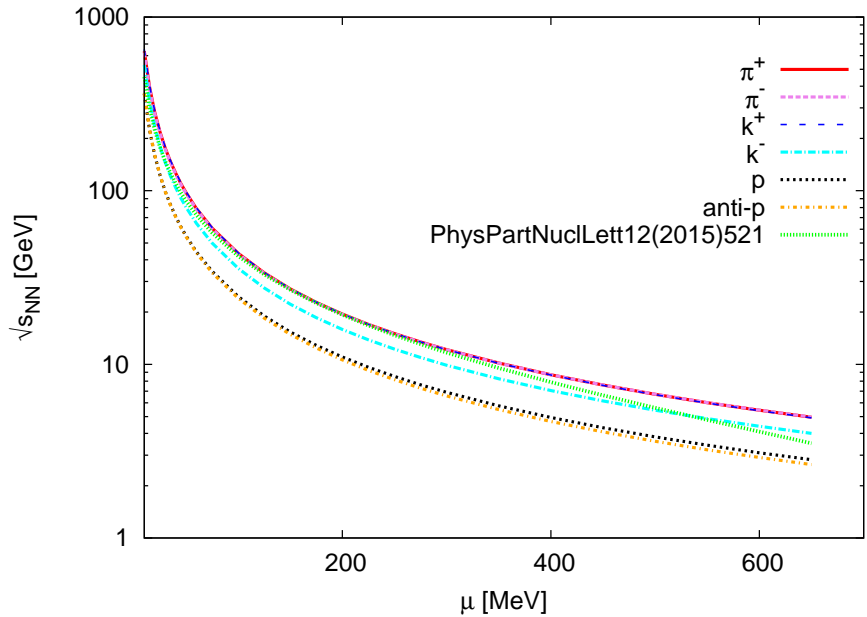}
\caption{In a semi-log plane, $\sqrt{s_{\mathtt{NN}}}~\mathtt{[GeV]}$ is depicted in dependence on $\mu~\mathtt{[MeV]}$ for $\pi^+$, $\pi^-$, $K^+$, $K^-$, $p$, $\bar{p}$ compared with the thermal model estimations  \cite{Tawfik:2013bza}.  \label{fig:1b}}
\end{figure}

The generic relation between beam energy and rapidity can be parameterized as follows.
\bea
\sqrt{s_{\mathtt{NN}}} = c\,  y^d,  \label{eq:sqrtsy}
\eea
where $c$ and $d$ are constants depending on the type of the particle. For $\pi^+$:
$c               = 119990         \pm 1217$ GeV, 
$d               = -2.322      \pm 0.003 $. 
For $\pi^-$:
$c               = 115433         \pm 1196$ GeV, 
$d               = -2.313      \pm 0.003.$ 
For $K^+$:
$c               = 103306         \pm 180.8$ GeV,    
$d               = -2.337        \pm 0.001$.  
For $K^-$:
$c               = 103323       \pm 180.8$ GeV, 
$d               = -2.337     \pm 0.001$. 
For $p$:
$c               = 64699.7        \pm 1955$ GeV,     
$d              = -2.30992       \pm 0.0116$. 
For $\bar{p}$:
$c               = 74134.7        \pm 2574$ GeV, 
$d               = -2.356       \pm 0.013$. 
When substituting Eq. (\ref{eq:muy}) into Eq. (\ref{eq:sqrtsy}) we get
\bea
\sqrt{s_{\mathtt{NN}}}~\mathtt{[GeV]} &=& c \left(\frac{\mu~\mathtt{[MeV]}-a}{b}\right)^{d/2}, \label{eq:sqrtsy2}
\eea
where the exponent $d/2$, which apparently merely depends on the type of particles, can be approximated as $-1.165$. 

Fig. \ref{fig:1b} depicts  the results corresponding to $\pi^+$, $\pi^-$, $K^+$, $K^-$, $p$, $\bar{p}$, Eq. (\ref{eq:muy}). The double-dotted curve gives the calculations based on thermal model  \cite{Tawfik:2013bza}. We find that pion and kaon have almost an identical energy-dependence, while proton and anti-proton have a slightly rapid energy-dependence. Nevertheless, the differences between both sets of calculations could be neglected. Firstly, the present study deals with a {\it generic} chemical potential, while the expression based on thermal models \cite{Tawfik:2014eba,Tawfik:2013bza} deals with the baryon type, only. Secondly, the present study focuses on single particle and its rapidity in additional to the transverse components of mass and momentum, while ref. \cite{Tawfik:2013bza} is based in bulk themrodynamics of an ensemble of an ideal gas with various components, hadron resonances \cite{Tawfik:2014eba}. When comparing both expressions, we observe that the resulting curves have almost the same energy-dependence, at least quantitatively. As elaborated, the slight difference can be understood due to source and nature of the chemical potential. 

To summarize this point, we highlight once again that we are introducing a {\it generic} chemical potential. The produced particles taken into consideration account for both baryon and strangeness chemical potentials, only. This might be emphasized from the slight different between Kaons and protons, Fig. \ref{fig:1b} and Eq. (\ref{eq:sqrtsy2}). 

When reviewing the fit parameters obtained from Eq. (\ref{eq:sqrtsy2}), there is almost no difference between particle and anti-particle. This might be understood from the small differences in the measured quantities, $m_\perp$ GeV, $p_{\perp}$ GeV, and $d^2 N/(2 \pi p_{\perp} dp_{\perp} dy)~$GeV$^{-2}$ \cite{Adamczyk:2017iwn}. On the other hand, this result agrees well with the statistical thermal treatment, where the absolute chemical potential are taken identical for particle and anti-particle.

\subsection{Verification}
\label{Impl}

\begin{figure}[!htb]
\includegraphics[width=5.cm]{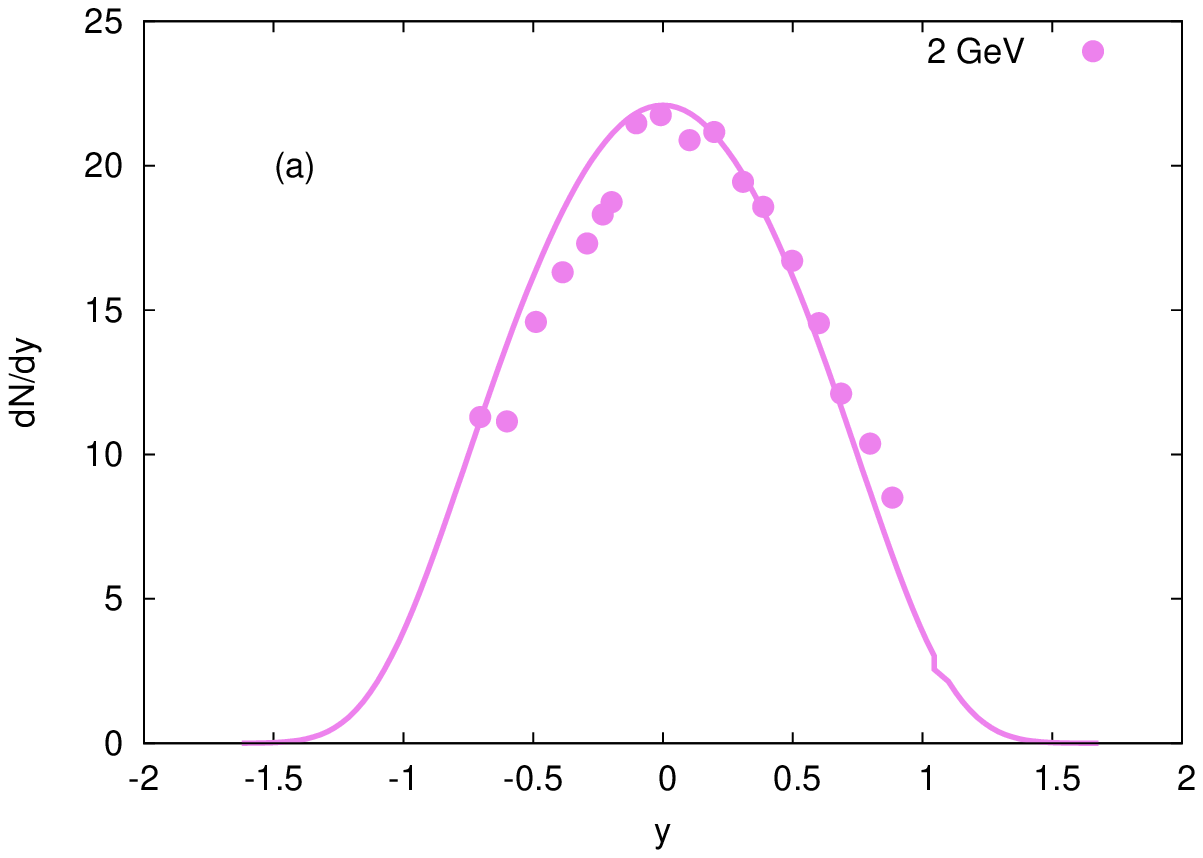}
\includegraphics[width=5.cm]{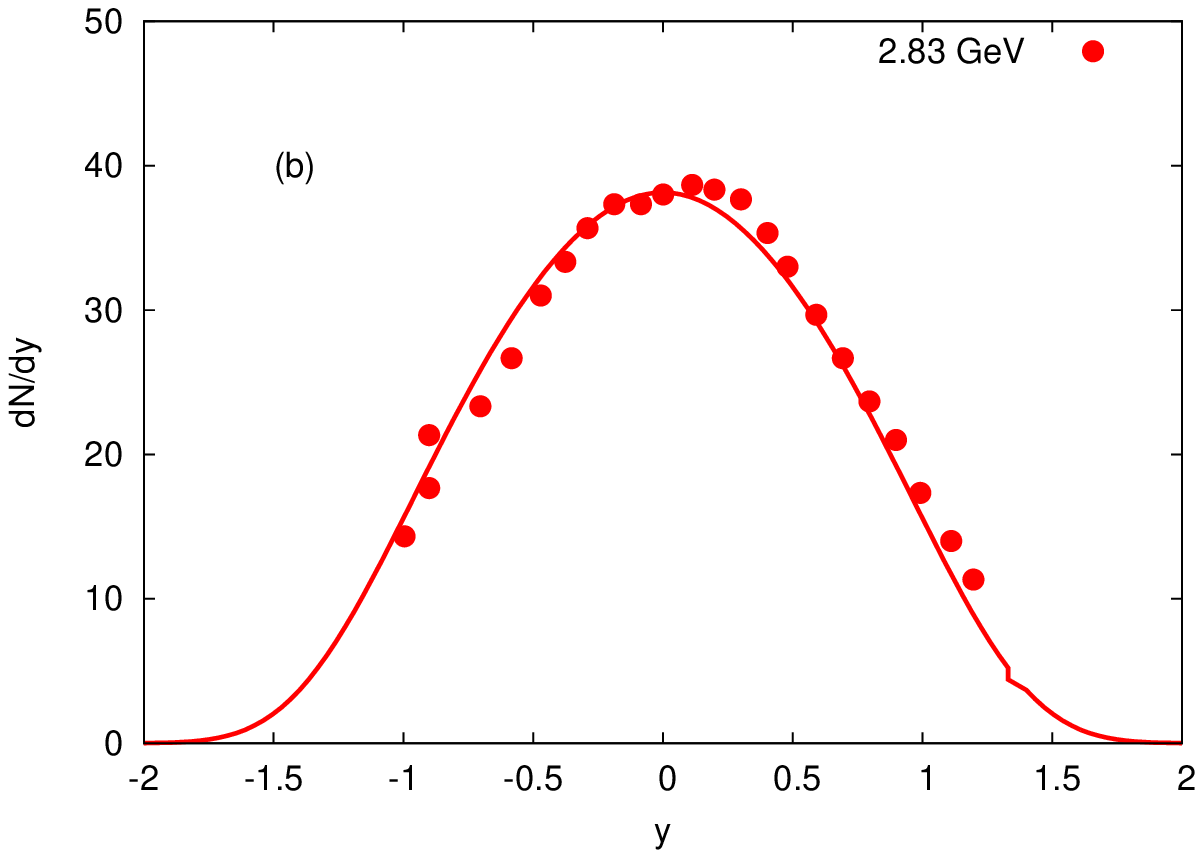}
\includegraphics[width=5.cm]{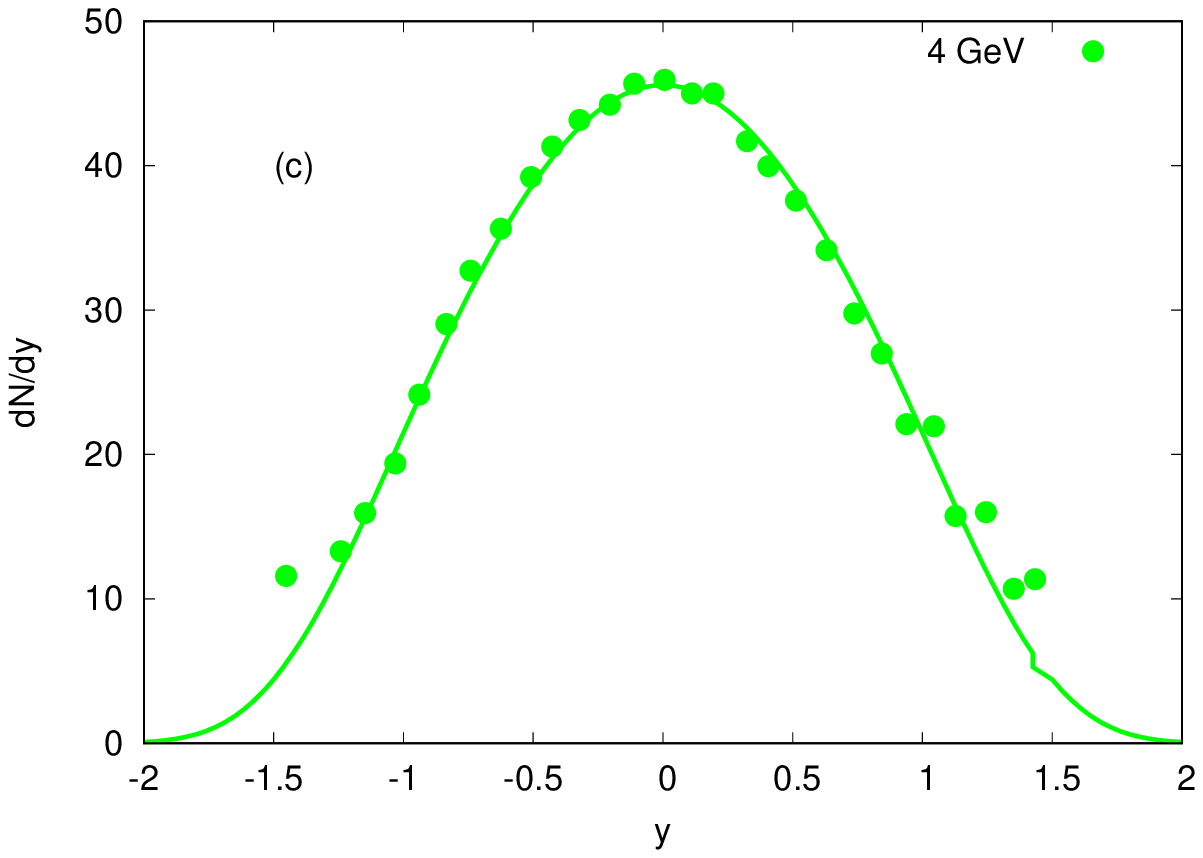}
\includegraphics[width=5.cm]{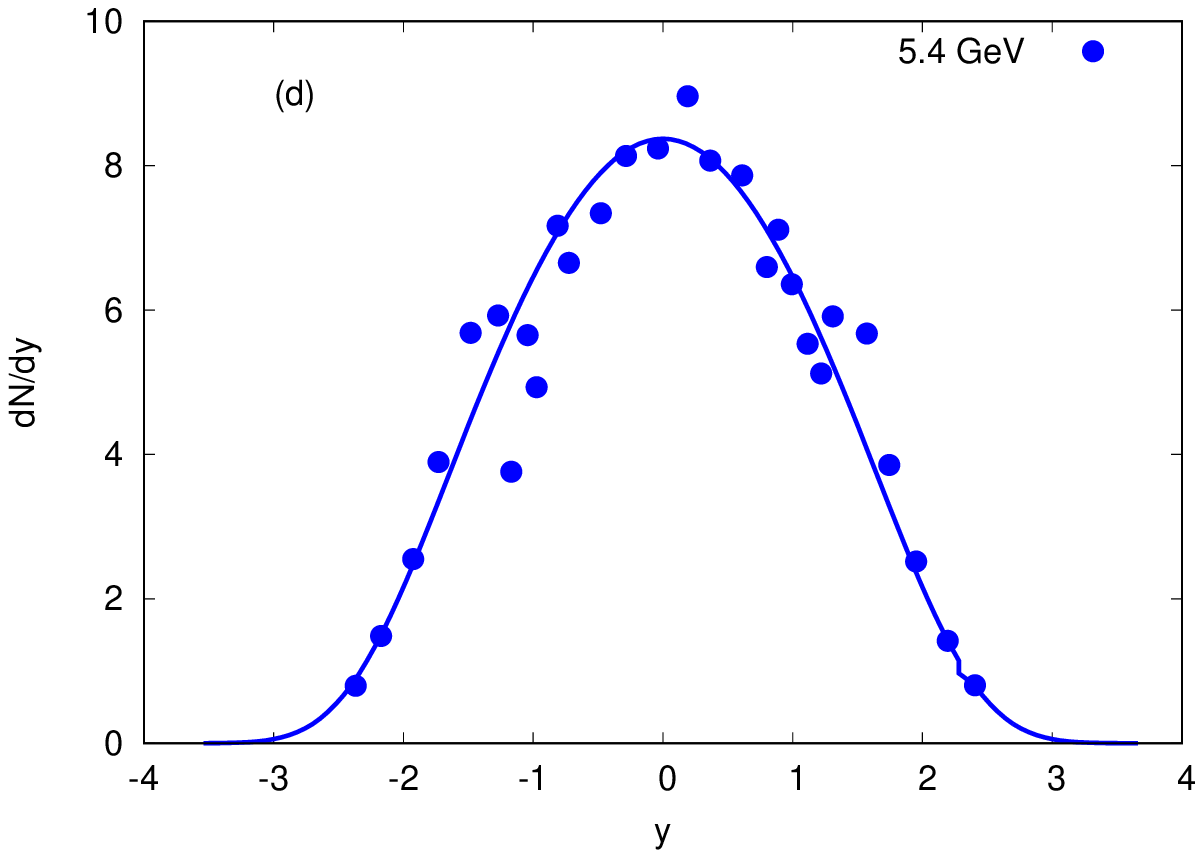}
\includegraphics[width=5.cm]{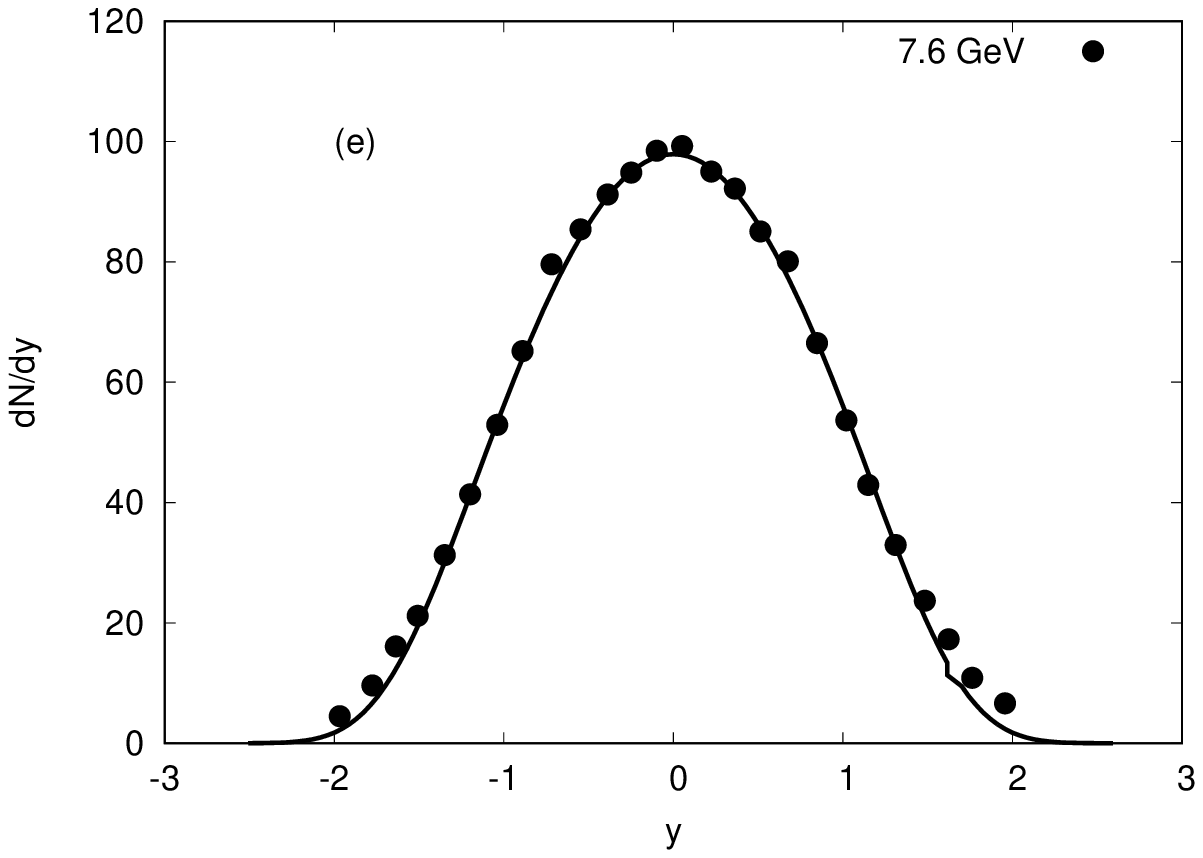}
\includegraphics[width=5.cm]{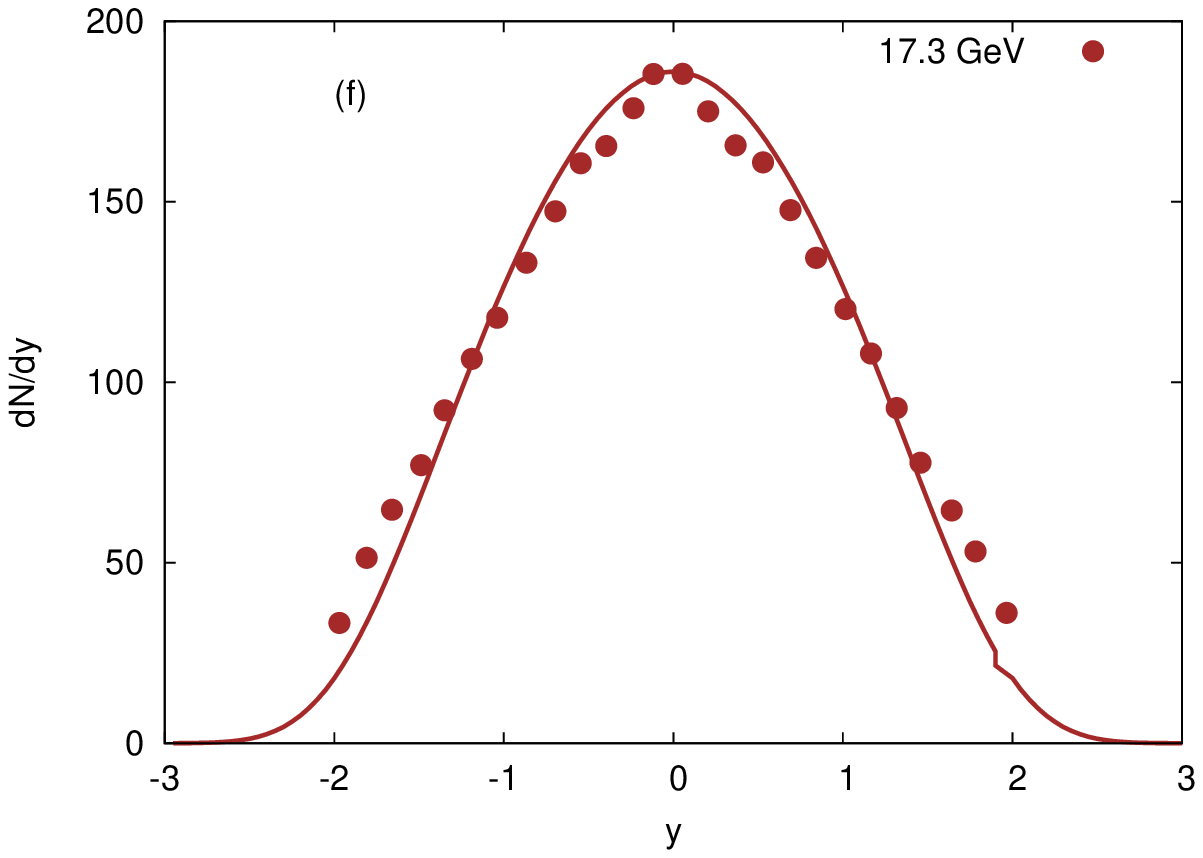}
\includegraphics[width=5.cm]{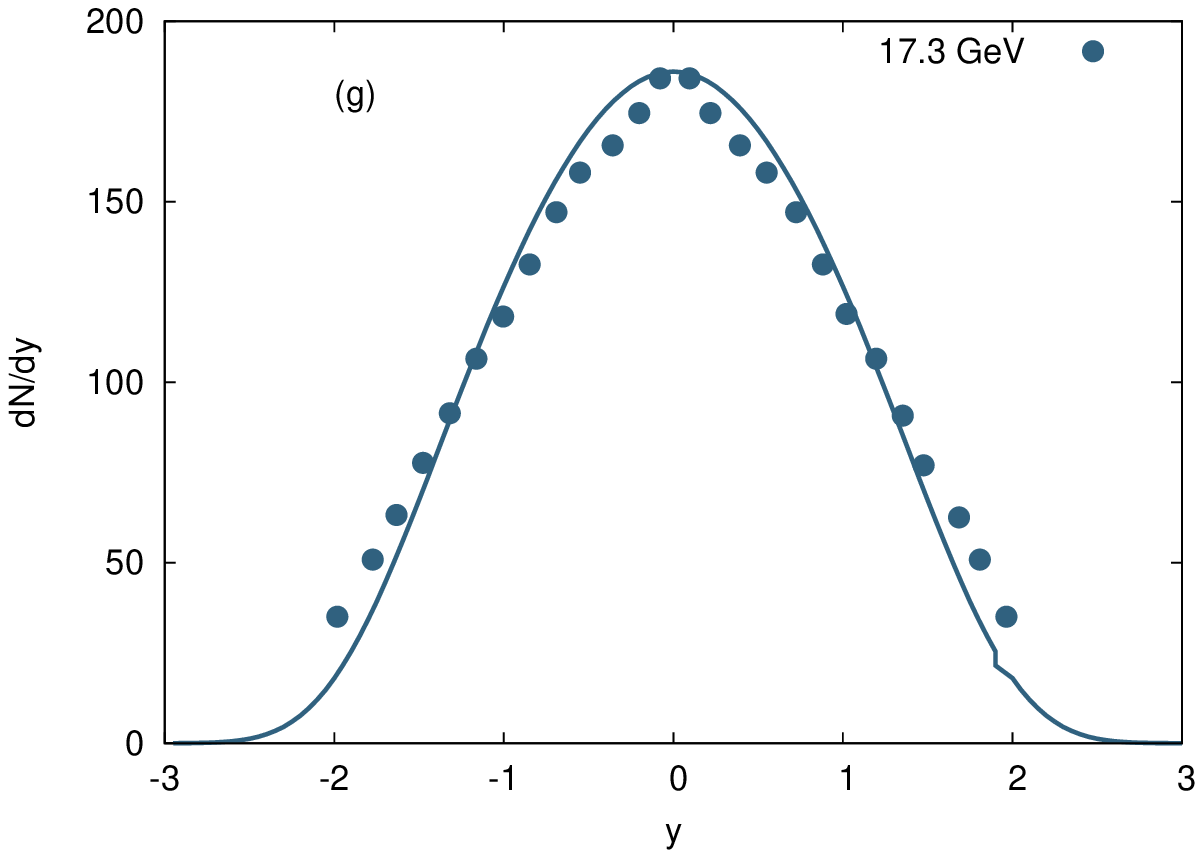}
\includegraphics[width=5.cm]{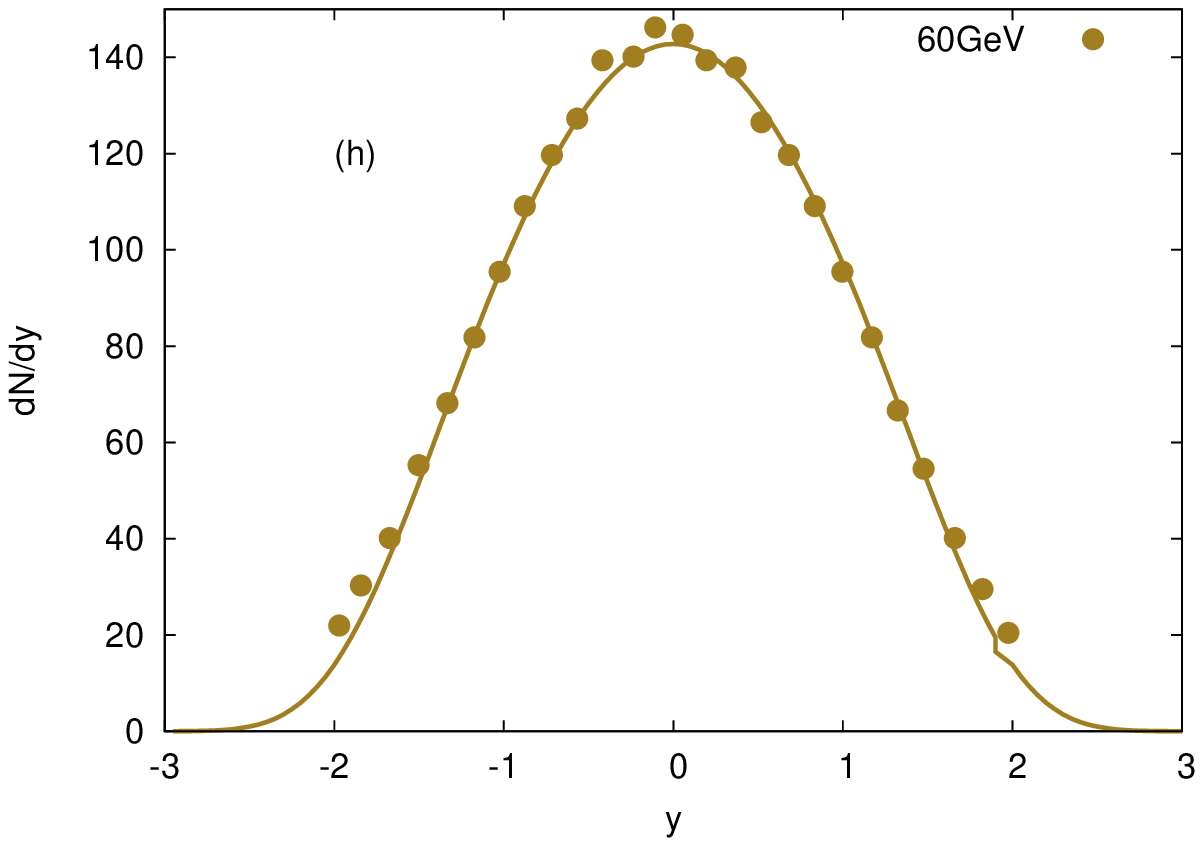}
\includegraphics[width=5.cm]{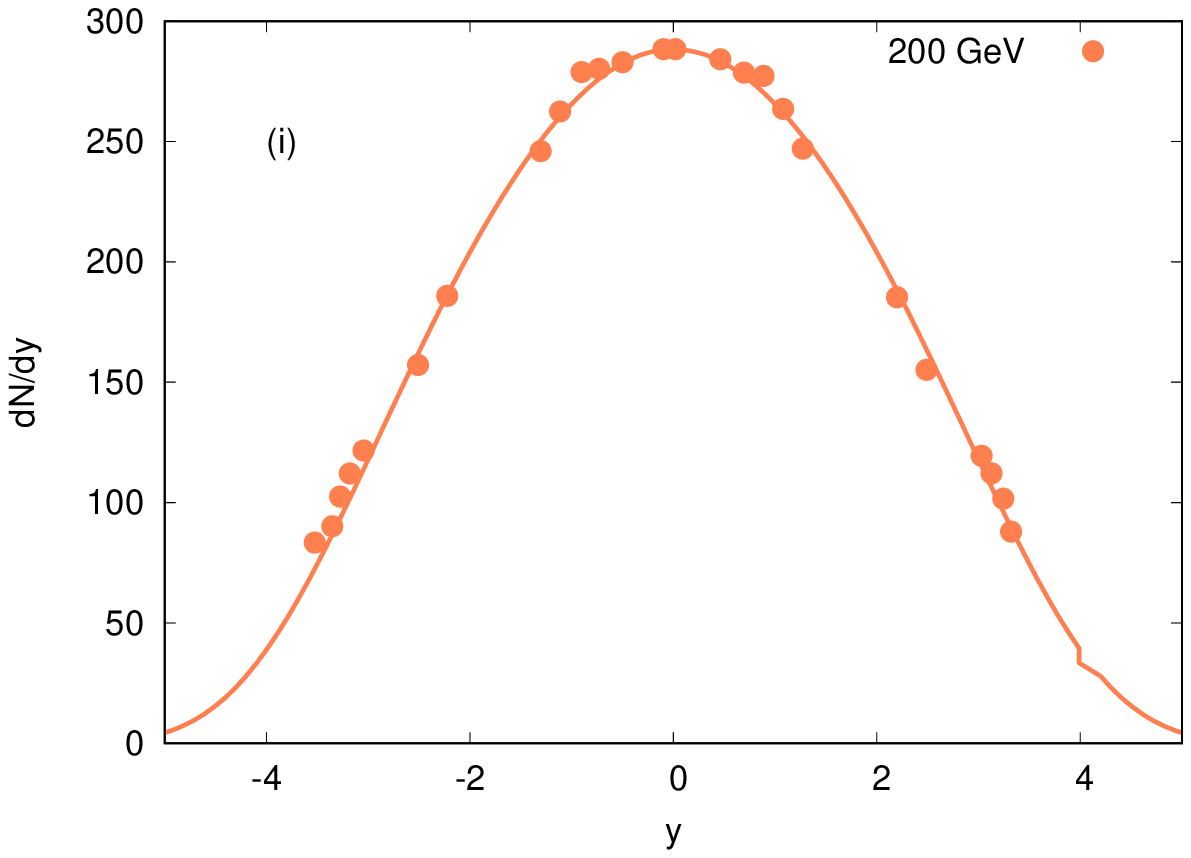}
\includegraphics[width=5.cm]{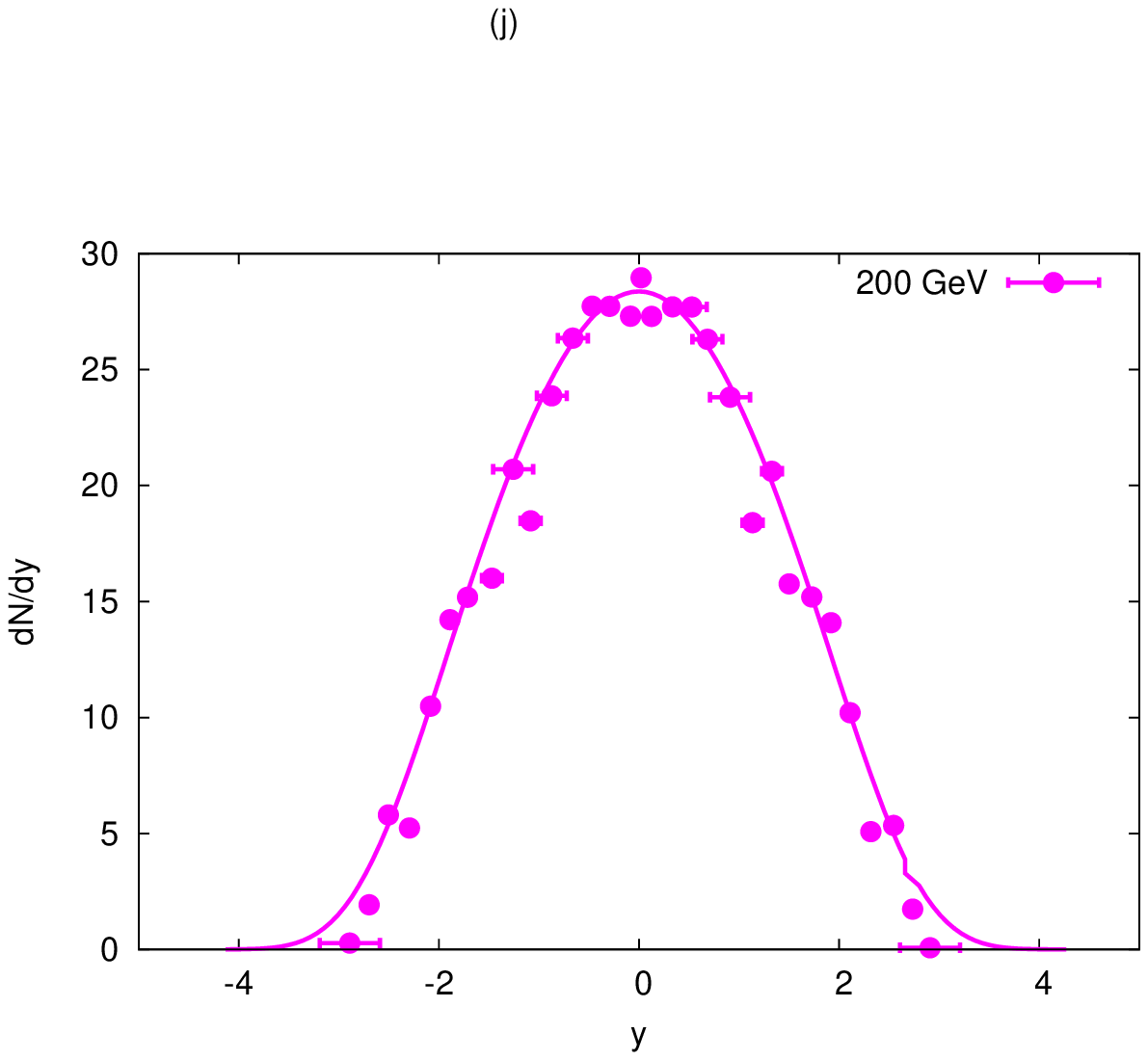}
\caption{The rapidity distribution $dN/dy$ is given as function of $y$ for $\pi^+$ (almost the same for $\pi^-$) measured in most central collisions, at energies ranging from $2~$GeV to $200~$GeV  \cite{Bearden:2003fw,Bearden:2004yx,Baechler:1994qx,Bartke:1990cn,Back:2000gw}. The experimental results are depicted as closed symbols while our calculations based on Eq. (\ref{eq:10}) are gien as solid curves.\label{fig:2} }
\end{figure}

\begin{figure}[!htb]
\includegraphics[width=5.cm]{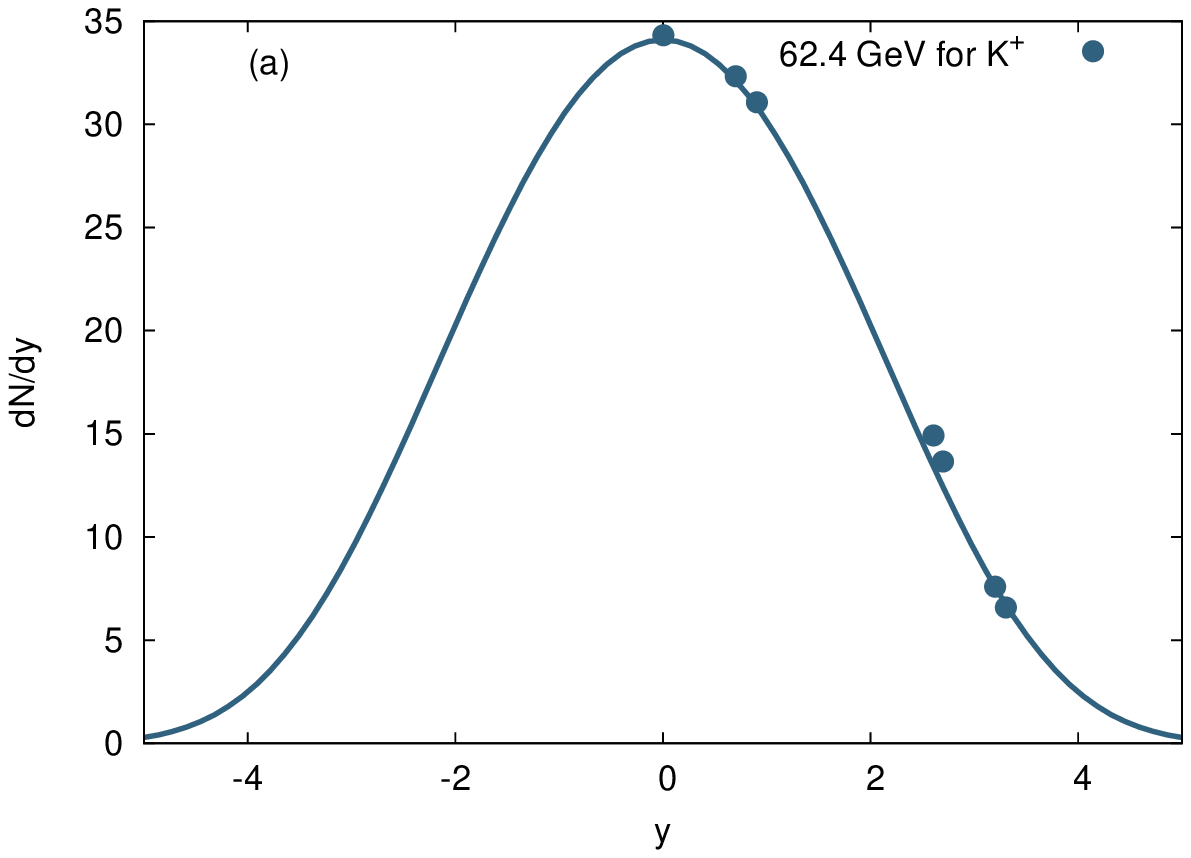}
\includegraphics[width=5.cm]{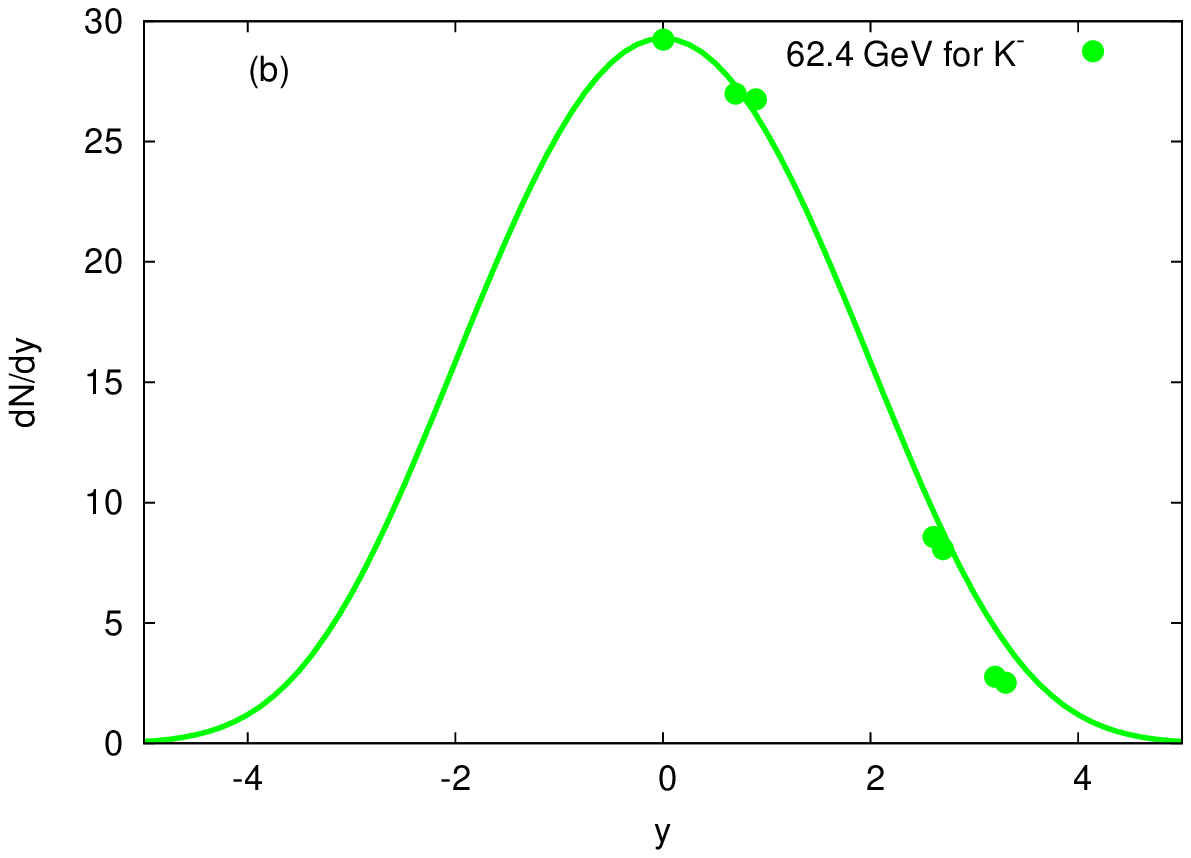}
\includegraphics[width=5.cm]{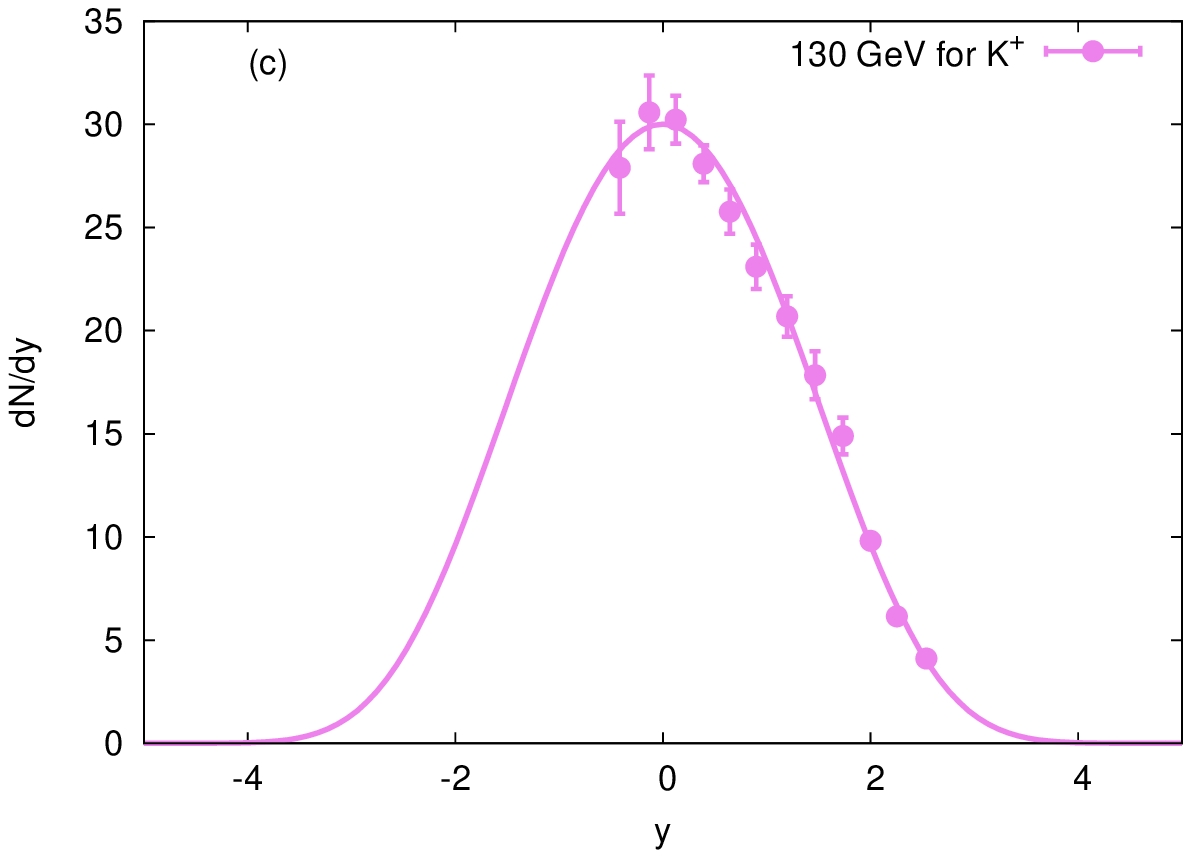}
\includegraphics[width=5.cm]{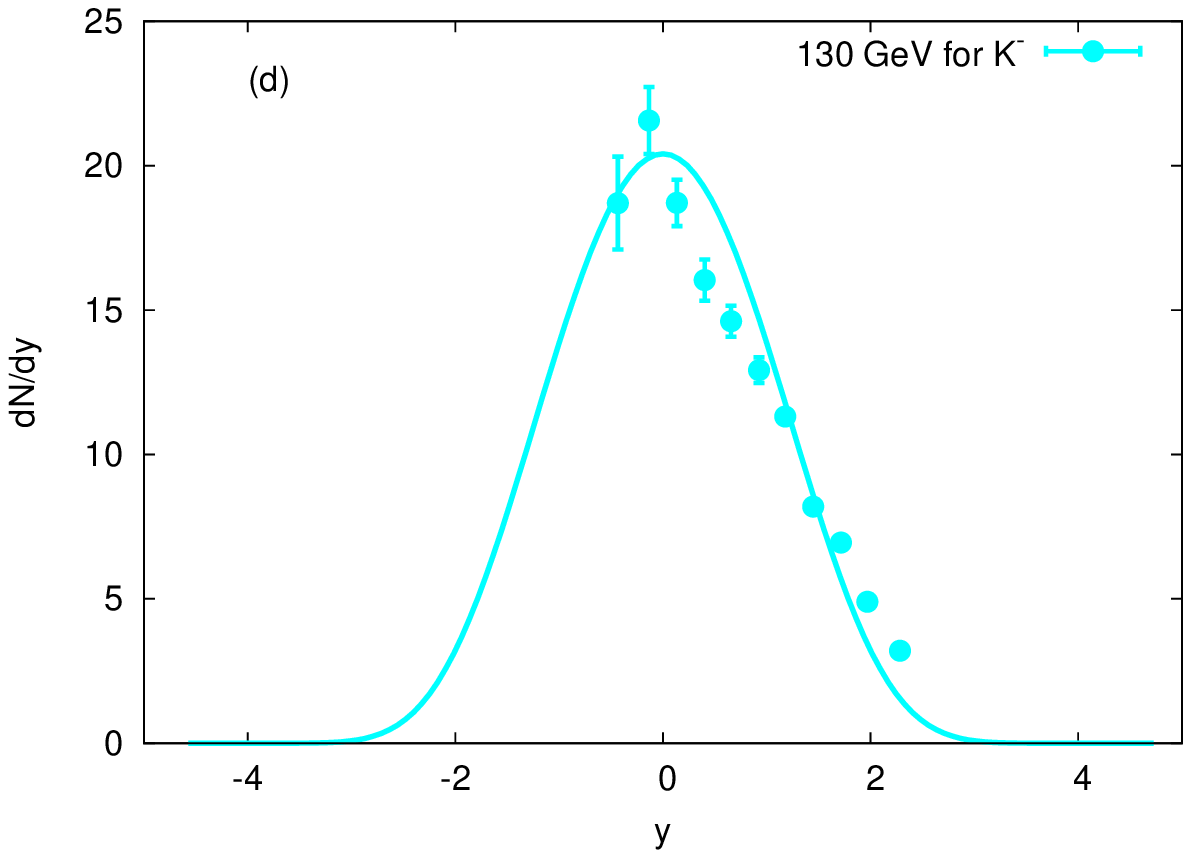}
\includegraphics[width=5.cm]{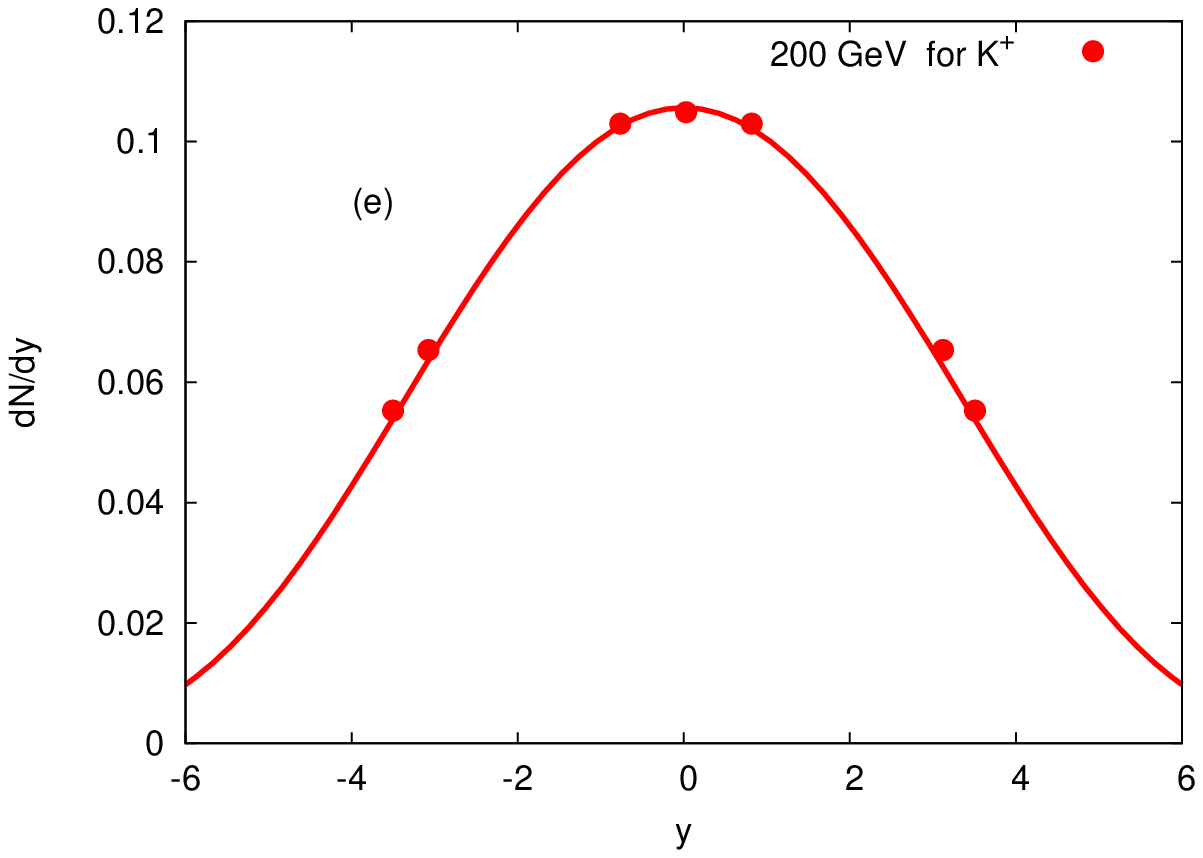}
\includegraphics[width=5.cm]{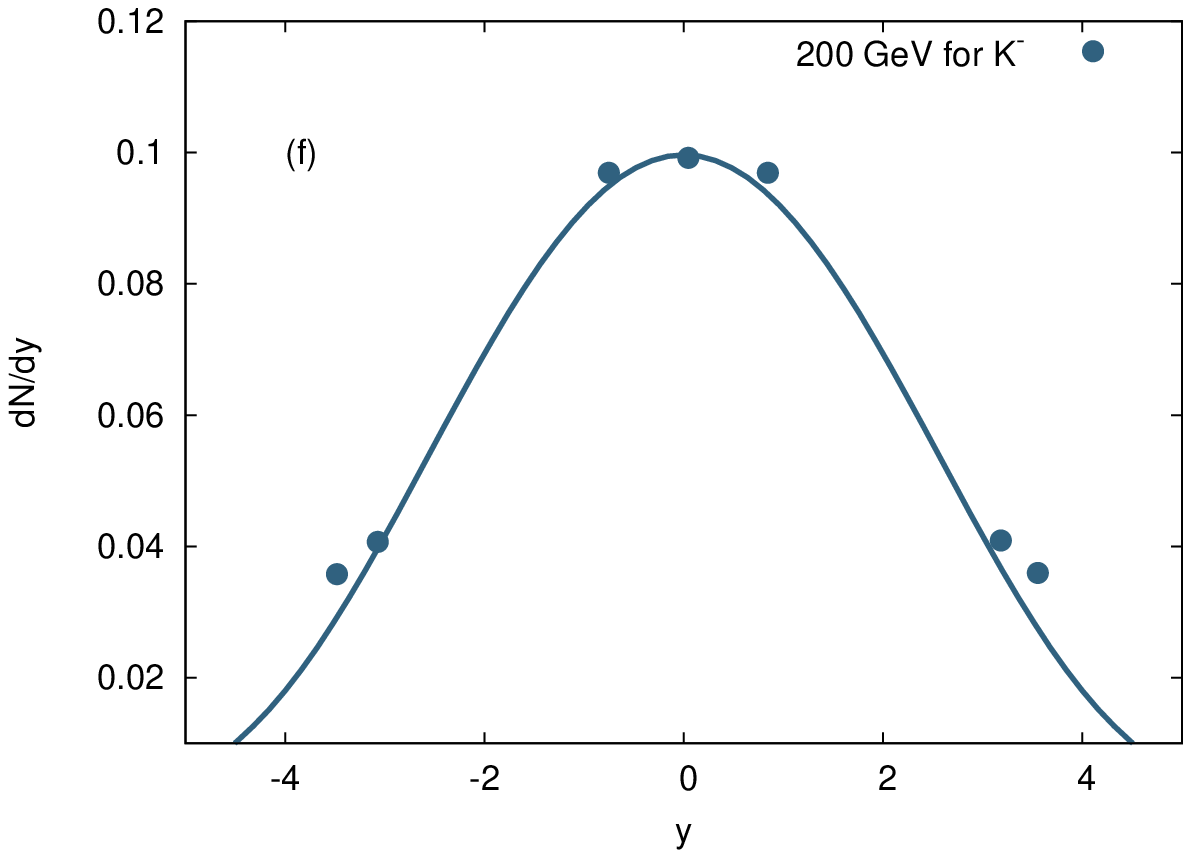}
\caption{The rapidity density distribution $dN/dy$ is given as function of rapidity $y$ for $K^+$  and $K^-$ at $\sqrt{s_{\mathtt{NN}}}=62.4, 130,$ and $200~$GeV \cite{Arsene:2009jg,Lin:2000cx,Bearden:2003fw,Bearden:2004yx}. The experimental results are depicted as closed symbols while our calculations as solid curves. 
\label{fig:3}}
\end{figure}

\begin{figure}[!htb]
\includegraphics[width=5.cm]{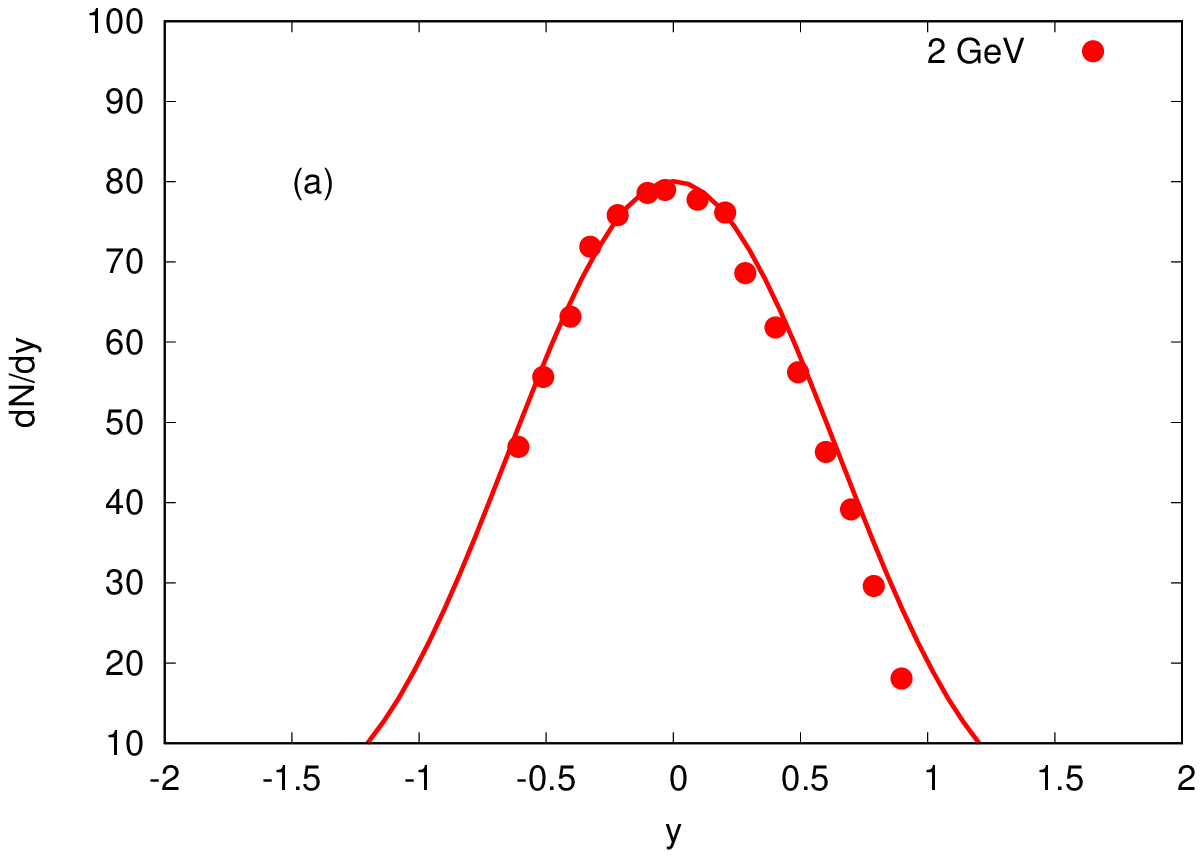}
\includegraphics[width=5.cm]{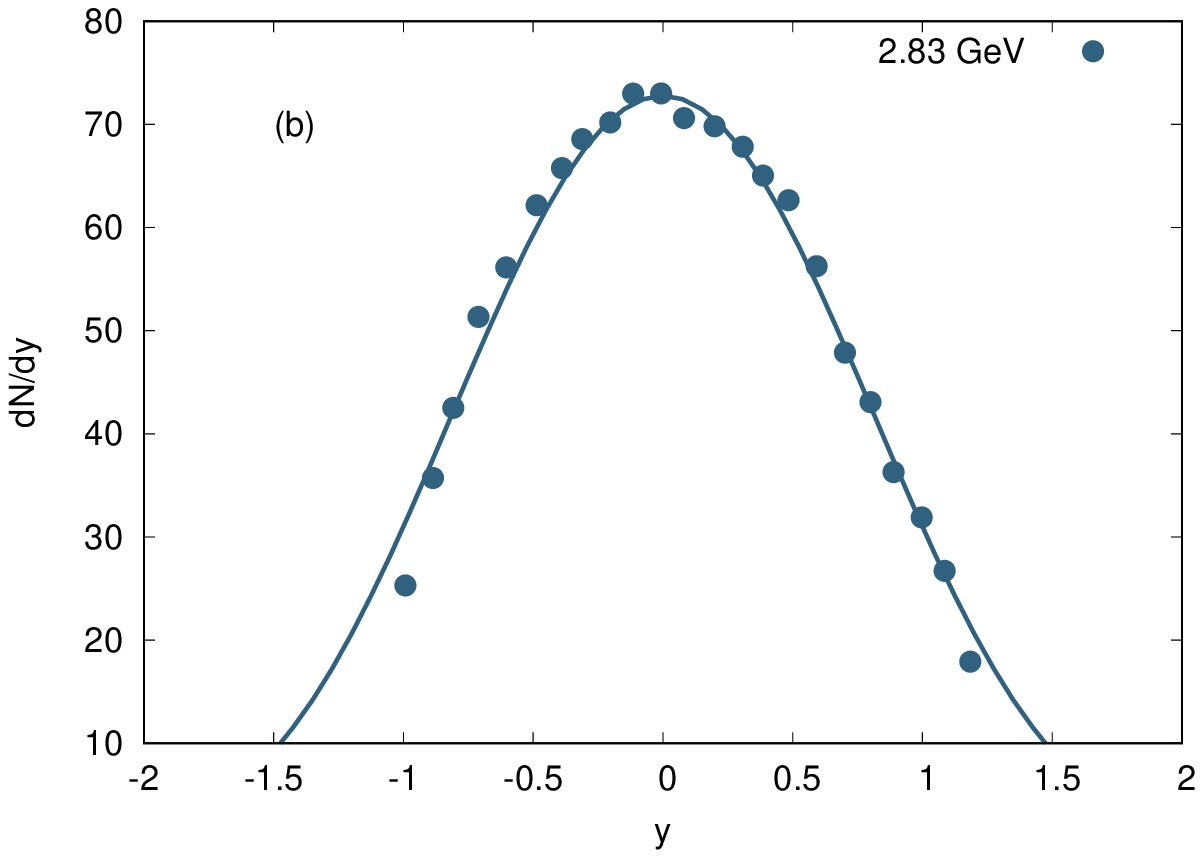}
\includegraphics[width=5.cm]{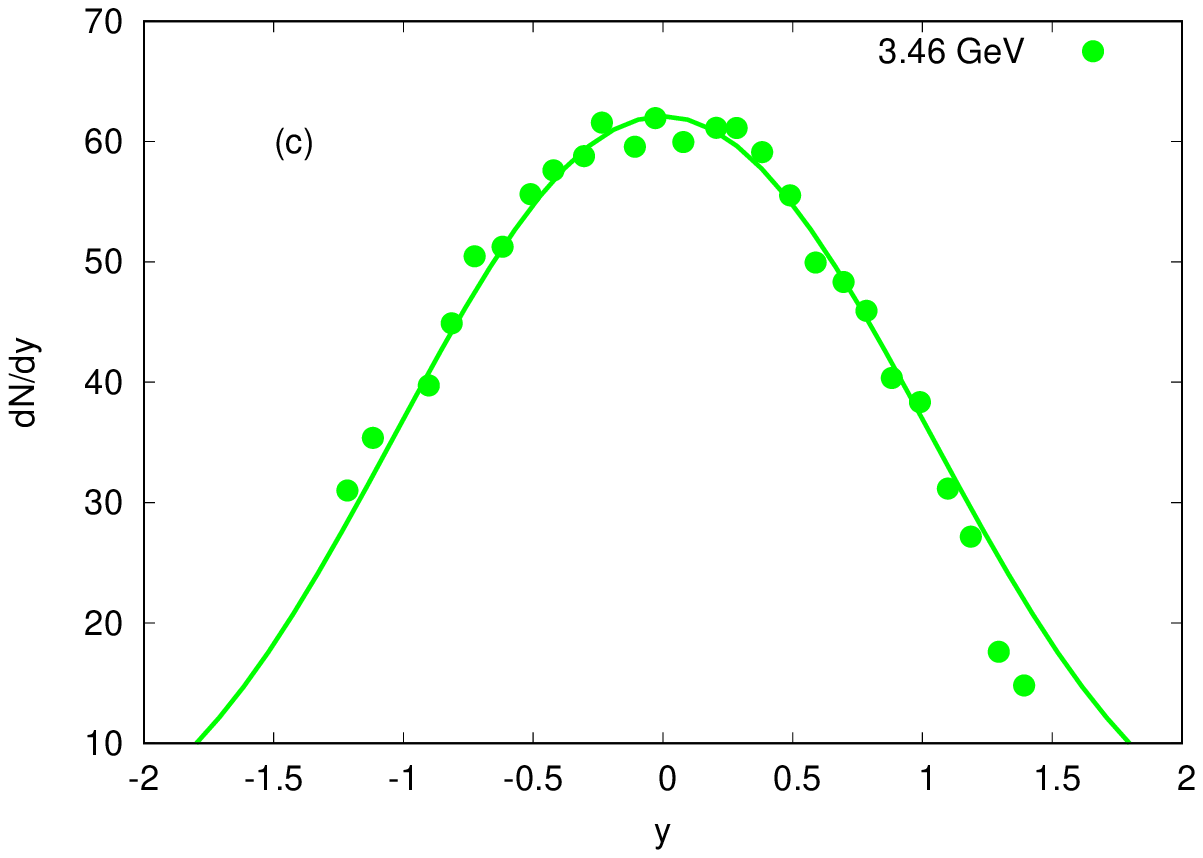}
\includegraphics[width=5.cm]{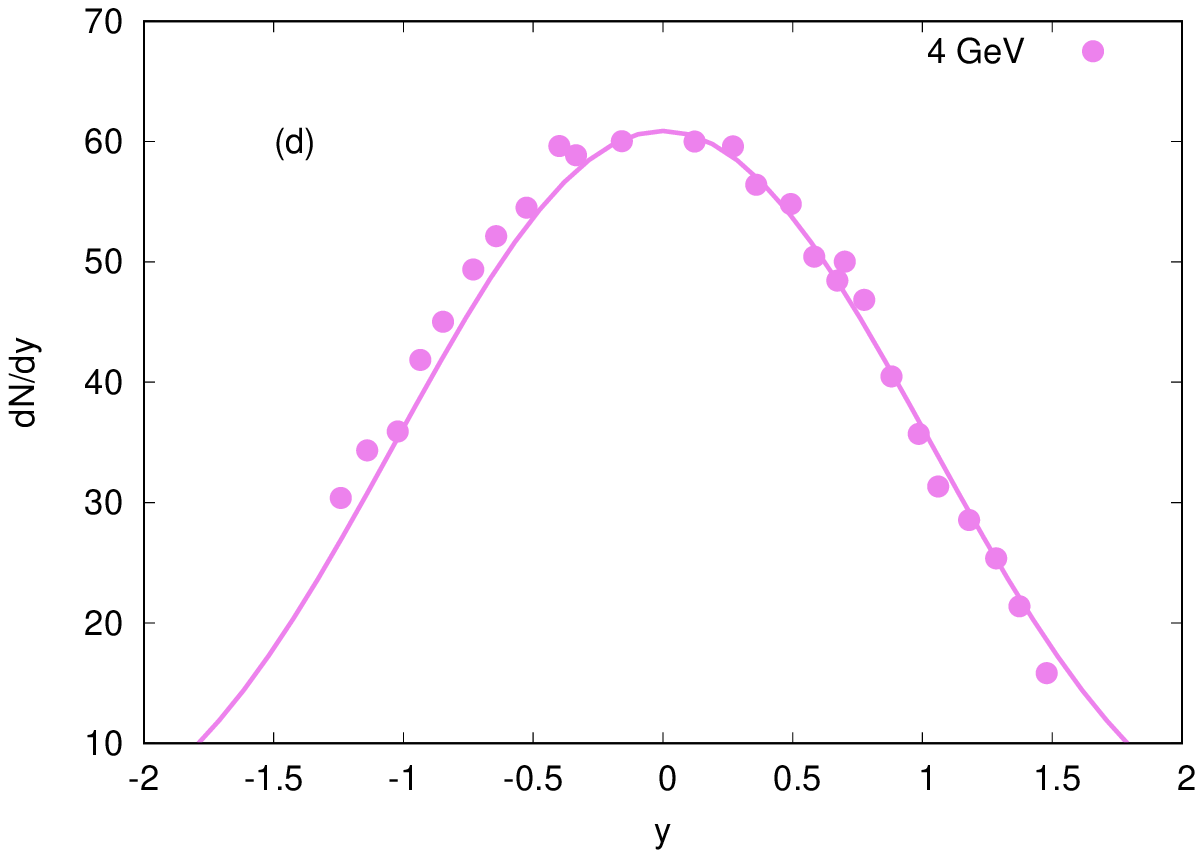}
\includegraphics[width=5.cm]{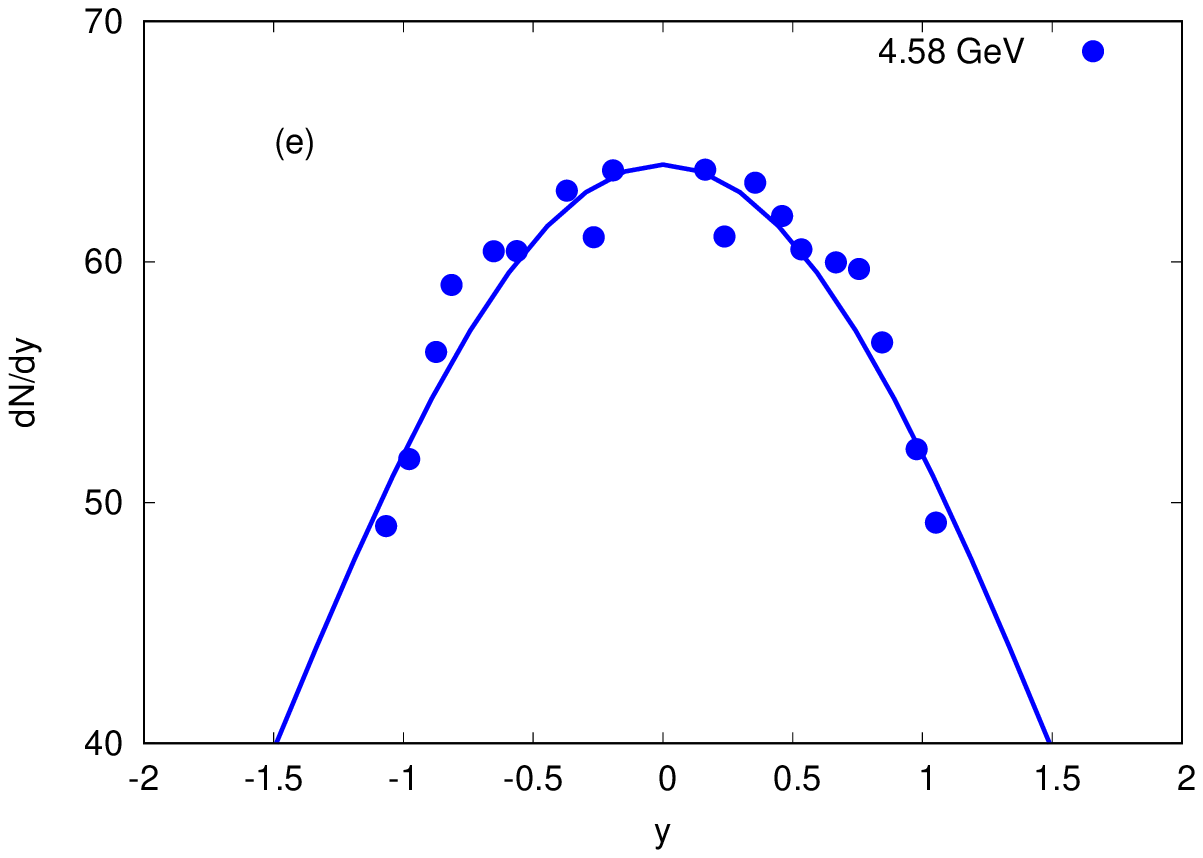}
\includegraphics[width=5.cm]{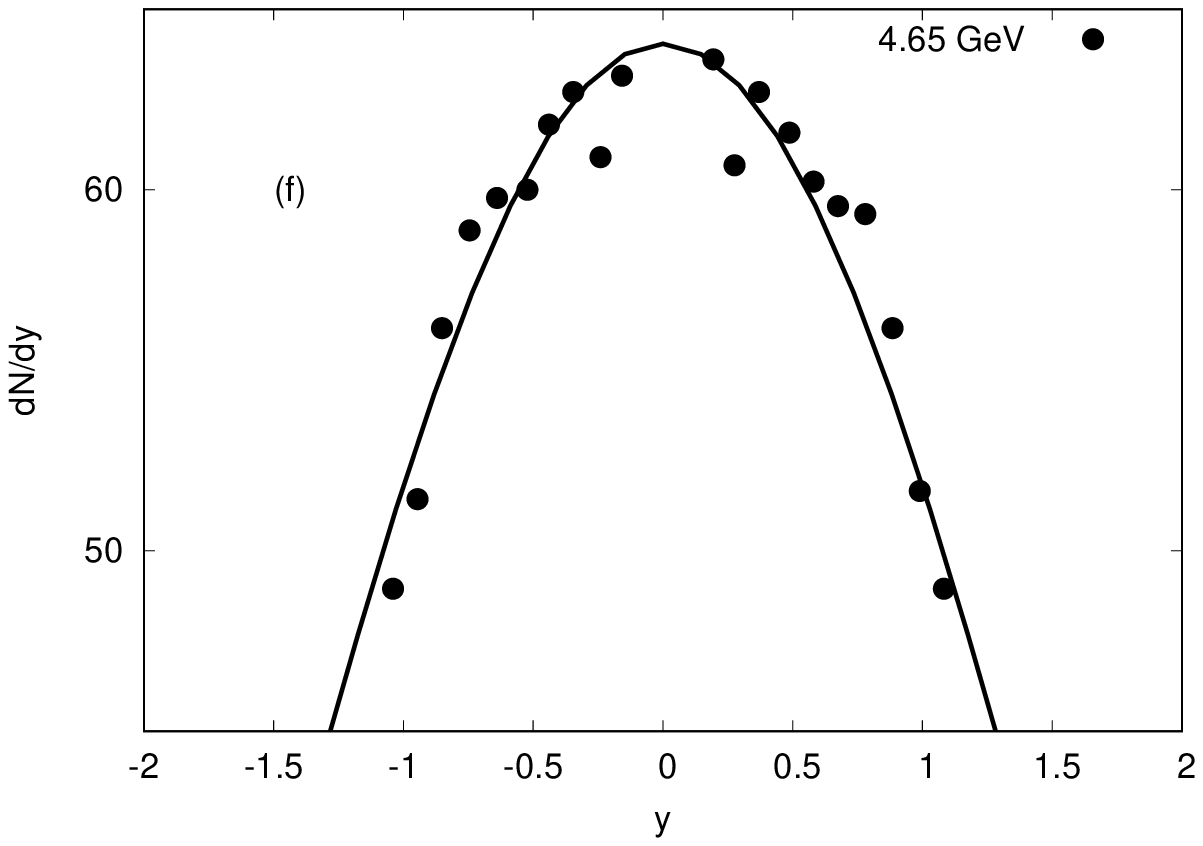}
\includegraphics[width=5.cm]{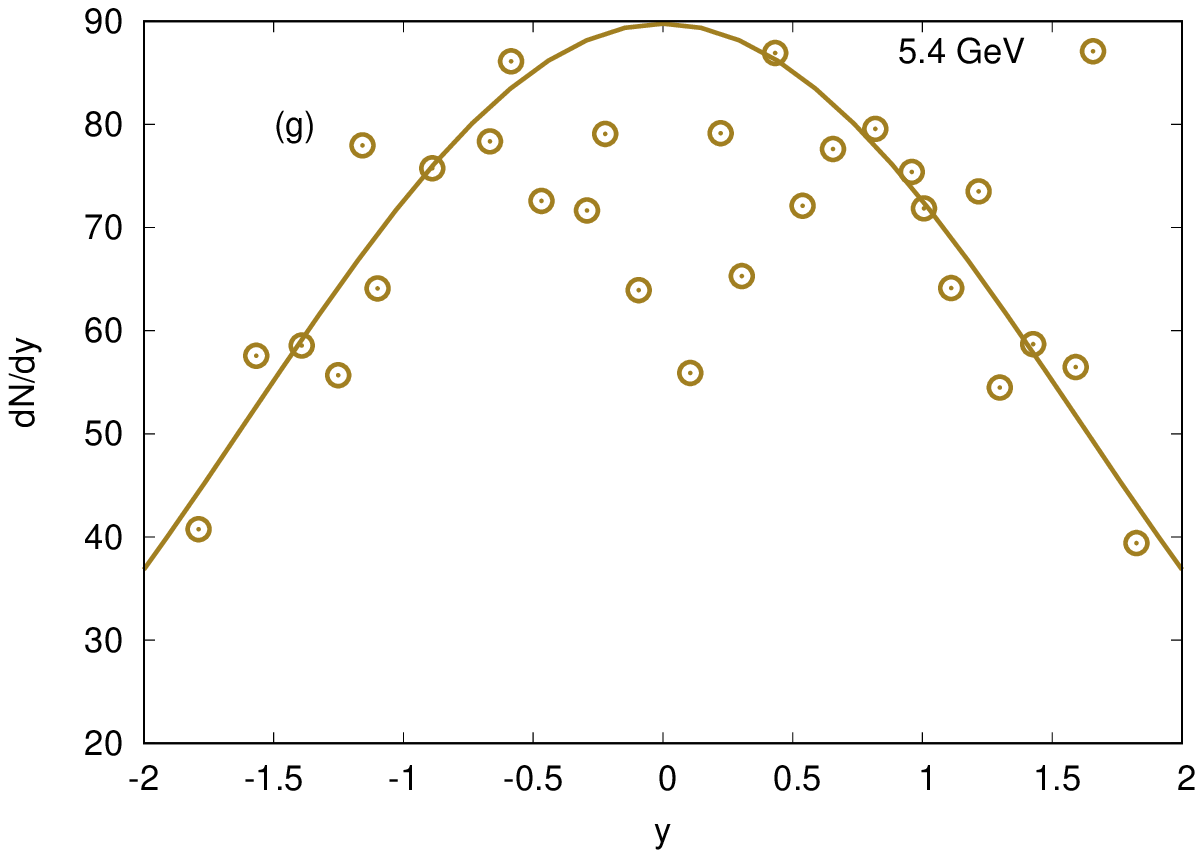}
\includegraphics[width=5.cm]{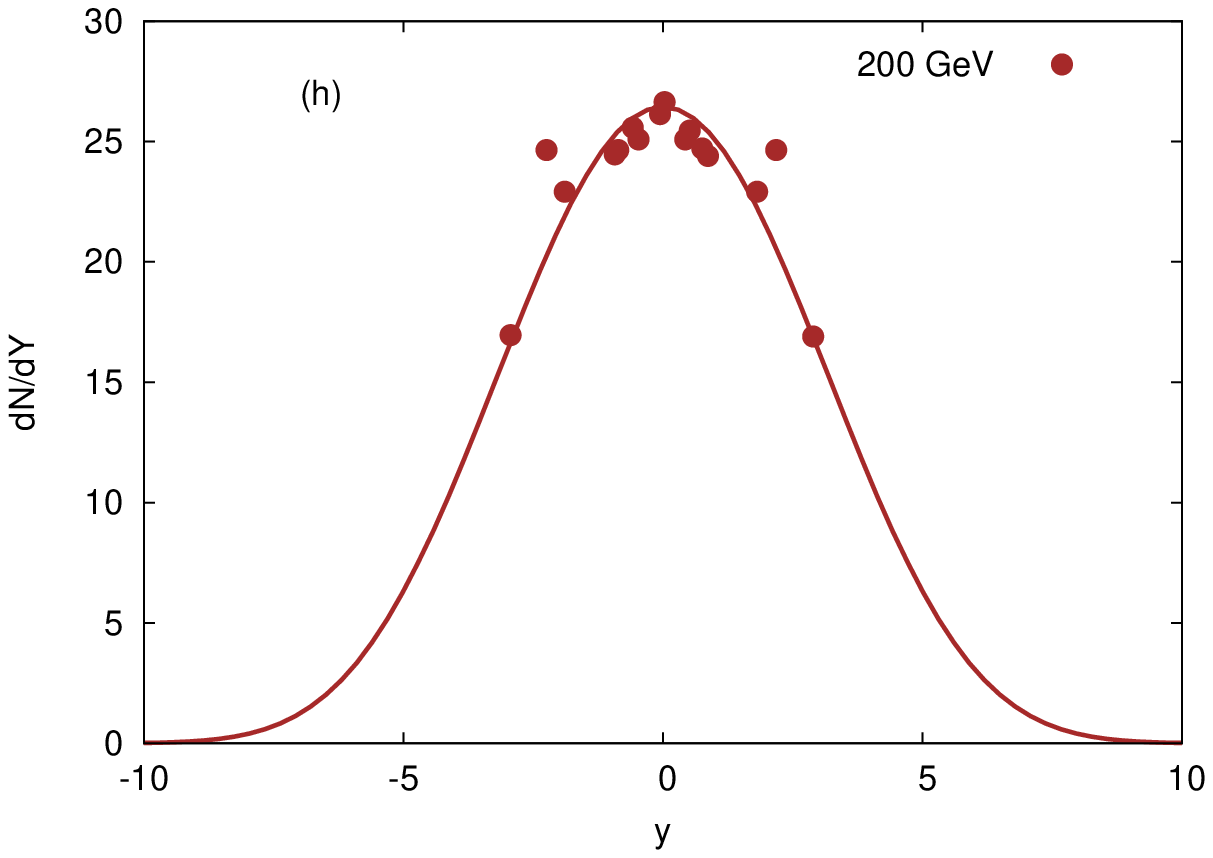}
\caption{The same as in Fig. \ref{fig:2} but for protons and antiprotons (almost the same for anti-proton).
\label{fig:4}}
\end{figure}

One of the validity checks was elaborated in the previous section. To this end, we have started with Eq. (\ref{eq:muy}), in which the chemical potential $\mu$ is substituted by Eq. (\ref{eq:sqrtsy}). The constants in Eq. (\ref{eq:sqrtsy}) are given in Tabs. \ref{Tab:1}-\ref{Tab:6}, which apparently vary from particle to particle. In other words, the dependence of $\mu$ on $y$ is simply given by the constants $a$ and $b$, which in turn depend on the particle of interest. Finally, we have illustrated Eq. (\ref{eq:sqrtsy2}) in Fig. \ref{fig:1b}. The results agree well the the thermal model calculations \cite{Tawfik:2013bza}.

In the present section, we discuss another validity check. We compare the experimental results on $dN/dy$ vs. $y$ with our calculations based on Eq. (\ref{eq:10}). As introduced in section \ref{subsec:thermal}, for these calculations, $\mu$ given in Eq. (\ref{eq:9}) plays an important role. Besides $\mu$, the corresponding $p_{\bot}$ and $m_{\bot}$ characterize the particle of interest. Confronting these calculations to the experimental $dN/dy$ vs. $y$, which is a very precise measurement proves or disproves the validity of the proposed approach. Should our approach for $\mu$, Eq. (\ref{eq:10}), is valid, the comparison, where no statistical fits are done, agrees well with the experimental results. 

Figure \ref{fig:2} shows the rapidity distribution $dN/dy$ for $\pi^+$  (almost the same for $\pi^-$) as functions of $y$ at energies ranging from $2~$GeV to $200~$GeV (symbols) \cite{Bearden:2003fw,Bearden:2004yx,Baechler:1994qx,Bartke:1990cn,Back:2000gw}. The calculations using Eqs. (\ref{eq:10}) are given as solid curves. It is apparent that there is an excellent agreement, at all energies.  

Figure \ref{fig:3} depicts another comparison for $K^+$ and $K^-$, at energies ranging from $62.4~$GeV to $200~$GeV \cite{Arsene:2009jg,Lin:2000cx,Bearden:2003fw,Bearden:2004yx}. Again, our results, Eq. (\ref{eq:10}), agrees well with the experimental results.

Figure \ref{fig:4} presents $dN/dy$ for proton (almost the same for anti-proton), at different energies \cite{Bachler:1999hu,Bearden:2003fw,Biedron:2006vf}. It is also apparent that Eq. (\ref{eq:10}) is excellently successful in determining the measured $dN/dy$ vs. $y$ for this baryon particle, at different energies.

As shown in figures \ref{fig:2}, \ref{fig:3}, \ref{fig:4}, there is a excellent agreement between the experimental results and our calculations based on Eq. (\ref{eq:10}) for all particles. at various energies. The chemical potentials $\mu$, Eq. (\ref{eq:9}), besides other empirical inputs are crucial. Accordingly, we conclude that our approaches, Eqs. (\ref{eq:9}) and (\ref{eq:10}), 

It should be remarked that the last plot in Fig. \ref{fig:4} shows that the experimental results seem to have two peaks, while the curve, our estimation, has just one. This is the only disagreement. A similar case would be also seen in the last plot of Fig. \ref{fig:2}. To fit with scope of the present script, we highlight that the curve is solely obtained by using Eq. \ref{eq:10}, e.g. no statistical fits have been conducted. With these regards, one would have noticed that both plots, in which this discrepancy appears, depict the results at the highest energies available. This might manifest remarkable signatures to be analyzed in a future study. A more plausible explanation is still missing.

\section{Conclusions}
\label{sec:cncl}

Despite the restrictions put forward by the availability of experimental results on $p_{\bot}$ and $d^2 N/(2 \pi p_{\bot} dp_{\bot} dy)$ or accessibility of such results, which enforced us to compare few results measured at different energies, we believe that this is an excellent opportunity to check the ability of our proposed approach in describing excellently all such experimental results and accordingly to show that the universal dependence on $\mu$ on $y$ works, well. We have restricted the calculations to most-central collisions. This vary from measurement to measurement and of course from experiment to experiment. It can be $0-5\%$ centrality or $0-10\%$ centrality. We focused on the narrower ones. 

Based on the experimental results  on $p_{\bot}$ and $d^2 N/(2 \pi p_{\bot} dp_{\bot} dy)$ for the well-identified produced particles, pions, kaons and protons and their antiparticles, the corresponding chemical potentials could be estimated in dependence on rapidity. By doing this, it was possible to derive a universal approach relating these two quantities with each other regardless type of the produced particles and energy and size of the colliding system $\mu=a+b y^2$, where $a$ and $b$ are constants. An excellent agreement was found also when comparing the resulting energy dependence $\sqrt{s_{\mathtt{NN}}}=c[(\mu-a)/b]^{d/2}$, where $c$ and $d$ are constants to be fixed, with the statistical thermal model. 

It wouldn't proper to make any clear statement about conservation. The reason is obviously based on procedure we are introducing.  The rough limitation is restricting our approach to single particles. On the other hand, the experimental inputs are apparently limited to the detector’s acceptance factor, and the ranges of rapidity and transverse momenta, etc. The assumption of an over all equilibrium seems should not contradict the proposed procedure which is restrictively focused on chemical potentials of single particles.  

The convincing agreement between our results for chemical potentials  obtained for mid-rapidity with the corresponding results of statistical models, Fig. \ref{fig:1b}, would encourage a new attempt to deduce the corresponding temperatures, the chemical freezeout ones, as well. This shall be subject of a future study. It should be highlighted that a constant temperature (freezeout temperature) is assumed in the present calculations, which is another limitation added to our approach.

Both expressions (\ref{eq:9}) and (\ref{eq:10}) are the main results of this script. Obviously, they are dealing with single particles, especially the ones which can be detected, precisely. Thus, the present approach is a kind of {\it local} chemical potential. Due the global equilibrium assured, our results apparently agree well with the energy dependence of $\mu$ which is based on statistical thermal approach for an indeal gas of hadron resonances \cite{Tawfik:2013bza,Andronic:2005yp}. Furthermore, the excellent reproduction of measured $dN/dy$ vs. $y$ for all particles at various energies gave a clear evidence supporting the validity of Eq. (\ref{eq:10}), in which $\mu$ estimated from Eq. (\ref{eq:9}) plays an essential role.

\bibliographystyle{aip}

\bibliography{mu_y8}

\end{document}